\documentclass[titlepage,12pt]{utarticle}
\usepackage{amssymb,amscd2000}
\usepackage{floatflt,graphicx} %useful for figures
\usepackage{dcpic,pictexwd} %really fancy commutative diagrams
\usepackage{color}
\usepackage[]{hyperref} %uncomment
\numberwithin{equation}{section}

\newcommand{\abs}[1]{\lvert #1\rvert}

\newcommand{\CE}{\sheaf{E}}

\newcommand{\category}[1]{\mathbb{#1}}
\newcommand{\complex}[1]{\sheaf{#1}^\bullet}

\newcommand{\Id}[2]{#1^{(#2)}}

\newcommand{\IdbarT}[2]{\overline{#1}^{\mathrm{T}(#2)}}
\newcommand{\mc}[1]{c_{#1}}
\newcommand{\mcbarT}[1]{\bar c^\mathrm{T}_{#1}}
\newcommand{\bbeta}{\boldsymbol{\beta}}
\newcommand{\bwp}{\boldsymbol{\wp}}
\newcommand{\bzeta}{\boldsymbol{\zeta}}
\newcommand{\ud}{\textrm{d}}
\newcommand{\uD}{\textrm{D}}
\newcommand{\uDbar}{\overline{\textrm{D}}}
\newcommand{\Lag}{\mathcal{L}}
\newcommand{\LKin}{\Lag_\textrm{kin}}
\newcommand{\LMul}{\Lag_\textrm{LM}}

\newcommand{\hc}{\textrm{h.c.}}

\newcommand{\bfield}[1]{#1}

\begin{document}
\preprint{
  UTTG--07--02\\
  {\tt hep-th/0206242}\\
}
\title{
  D-Brane Monodromies, Derived Categories and Boundary Linear Sigma Models
}
\author{Jacques Distler, Hans Jockers and Hyukjae Park
    \thanks{Work supported in part by NSF Grant PHY0071512
      and the Robert A.~Welch Foundation.
    }
}
\oneaddress{
      Theory Group, Physics Department\\
      University of Texas at Austin\\
      Austin, TX 78712 USA\\ {~}\\
      \email{distler@golem.ph.utexas.edu}
      \email{jockers@physics.utexas.edu}
      \email{hpark@zippy.ph.utexas.edu}
    }

\date{June 26, 2002}

\Abstract{
An important subclass of D-branes on  a Calabi-Yau manifold, $X$, are in 1-1 correspondence with objects in $D(X)$, the derived category of coherent sheaves on $X$. We study the action of the monodromies in K\"ahler moduli space on these D-branes. We refine and extend a conjecture of Kontsevich about the form of one of the generators of these monodromies (the monodromy about the ``conifold" locus) and show that one can do quite explicit calculations of the monodromy action in many examples. As one application, we verify a prediction of Mayr about the action of the monodromy about the Landau-Ginsburg locus of the quintic. Prompted by the result of this calculation, we propose a modification of the derived category which implements the physical requirement that the shift-by-6 functor should be the identity.
Boundary Linear $\sigma$-Models prove to be a very nice physical model of many of these derived category ideas, and we explain the correspondence between these two approaches
}

\maketitle

%%%%%%%%%%%%%%%%%%%%%%%%%%%%%%%%%%%%%%%%%%%%%%%%%%%%%%%%%%%%%%%%%%

\section{Introduction}\label{sec:intro}

Consider Type II string theory compactified on a manifold $X$. When the ``radius" of the compactification manifold, $X$, is large, D-branes can roughly be thought of as objects wrapped around cycles $C\in X$, equipped with some extra data (a vector bundle on $C$). A more precise characterization would consist of constructing the explicit boundary state, which is rarely possible, unless the conformal field theory corresponding to compactification on $X$ is rational.

If we set our sights lower, D-brane \emph{charge}, in this large-radius limit, is precisely characterized by the K-theory of $X$. Away from the large-radius limit, the group of D-brane charges is still a discrete abelian group, but its precise characterization is somewhat elusive. By way of creeping up on such a characterization, we can study the automorphisms of the group of D-brane charges (the K-theory of X) induced by traversing an incontractible loop in the moduli space of compactifications on $X$. Since such a loop may probe deep into the nongeometrical (``stringy") regime, we implicitly learn something about the stringy generalization of K-theory.

It is particularly interesting to study these monodromies in the case when $X$ is nonsimply-connected. Torsion phenomena distinguish K-theory from (co)homology, and hence can modify the naive picture above, even in the large-radius limit. In \cite{Distler-Brunner:Torsion,Distler-Brunner:TorsionII}, we studied D-branes on nonsimply-connected Calabi-Yau manifolds, $X$, and the monodromies of the group of D-brane charges as one moves in the moduli space of compactifications on $X$.

Here we would like to refine the picture presented in \cite{Distler-Brunner:Torsion,Distler-Brunner:TorsionII} by studying the action of the monodromies on the D-branes themselves, not just on the group of charges. To do this we make use of the proposal of \cite{DoFiRoI,DoFiRoII,Douglas:Categories} that a \emph{class} of D-branes on $X$ can be described as objects in $D(X)$, the (bounded) \emph{derived category} of coherent sheaves on $X$. It is certainly not true that all, or even most D-branes on $X$ can be described this way. The ones which \emph{can} are the $B$-type branes which are related to D-branes in the B-type topological string theory on $X$. 

These form a nice subclass of B-type branes which is, moreover, carried into itself by the action of the monodromies. As proposed by \cite{Kontsevich} and elaborated upon by \cite{Horja,HorjaII} and \cite{Thomas:MirrorBraid, Seidel-Thomas:Braid}, the monodromies act on these D-branes by \emph{auto-equivalences} of the derived category.

We will review Kontsevich's formula of the monodromy action on the derived category and will propose two modifications of his proposal. One will correct the grading, so that the D-branes which become massless at the (mirror of the) conifold point are invariant under the conifold monodromy
(as we expect to be the case, since there is a local field theory description of the physics near the conifold if one introduces these D-branes as fundamental fields in the action).

The second modification will be required to take account of the physics of \emph{nonsimply-connected} Calabi-Yau manifolds.

We will check these proposals by doing some explicit computations of D-brane monodromies on some simple Calabi-Yau manifolds. We will treat in detail the case of the quintic in $\BP^4$ and the orbifold of the quintic by a freely-acting $\BZ_5$ symmetry.

We will find explicit formul\ae\ for the monodromy action on wrapped D6-, D4- D2- and D0-branes. Among other things, we will verify a prediction of Mayr \cite{Mayr:McKayCorresp} about the orbit of the D6-brane under monodromy about the Landau-Ginsburg point.

However, in doing this explicit computation, we encounter a surprise, namely that the $5^{th}$ power of the Landau-Ginsburg monodromy is not the identity in the derived category. Rather, it is equal to the shift-by-12 functor. In \S\ref{sec:NewCategory}, we propose a modification of the derived category  which implements the physical requirement that the shift-by-6 functor is an isomorphism. In this modified category, $M_{LG}^5\simeq \Bid$. We propose this modified category as the correct category for B-type topological open strings. The new category has \emph{more} isomorphisms, and hence \emph{fewer} isomorphism classes (thus fewer D-branes). This resolves a puzzle \cite{Douglas:Categories} about the correspondence between the topological and the physical open string theory.

We may regard the derived category computations of the monodromies as mathematical ``predictions" for the behaviour of D-branes as we traverse an incontractible cycle in the moduli space. To get a physics ``verification" of these predictions, we need to specialize still further.

An interesting subclass of D-branes consists not of complexes of arbitrary coherent sheaves, but of objects in the derived category which are (quasi-isomorphic to) complexes of direct sums of \emph{invertible} sheaves (holomorphic line bundles). While this subclass doesn't have a nice categorical structure, it nonetheless may be realized using boundary linear $\sigma$-models. That means that we can use physical techniques to study the monodromies.

We introduce Boundary Linear $\sigma$-Models, in a formalism closely related to that of \cite{Hellerman-McGreevy:Toolshed}, and we show that there is a 1-1 correspondence between BL$\sigma$M data and objects in the derived category of the above form. Moreover, we argue that certain identifications on objects in the derived category, induced by the quasi-isomorphisms discussed in \S\ref{sec:Object simplification}, can be understood as deformations of the corresponding BL$\sigma$M. These deformations lead, in the infrared, to identical BCFTs.
Finally, we discuss to what extent we can recover the monodromies predicted by the derived category using the BL$\sigma$M.

\section{The Derived Category}
In this section, we review derived categories. The main purpose of
this section is
to set up our notation. By no means is it intended to give a
thorough review on the subject. For more rigorous treatment, see \cite{Gelfand-Manin}.

\subsection{Construction of the Derived Category}
We start with an Abelian category, $\category{A}$. Examples of abelian categories -- the ones which will be of most use to us -- are the category of abelian groups, the category of finite dimensional vector spaces, and the category of coherent sheaves on a manifold $X$. 

For a given Abelian category $\category{A}$, one can construct several 
categories induced from it. The first category we will consider is 
$\mathrm{Kom}(\category{A})$, \emph{the category of complexes over $\category{A}$}.
Objects of $\mathrm{Kom}(\category{A})$ are complexes in $\category{A}$ and morphisms are
chain maps. 
A \emph{complex} $\complex{E}$ in
$\category{A}$ is a sequence of objects and morphisms in $\category{A}$
\begin{equation}
\complex{E} : \cdots \xrightarrow{c_{n-1}} \sheaf{E}^n \xrightarrow{c_n}\sheaf{E}^{n+1}
\xrightarrow{c_{n+1}}\cdots
\end{equation}
with the property $c_n \circ c_{n-1} = 0$ for all $n$. In this paper, we will be exclusively interested in \emph{bounded} complexes, where the $\CE^n$ vanish except for a finite number of values of $n$. The corresponding category is usually denoted by $\mathrm{Kom}^b(\category{A})$, but we will, for ease of notation, drop the ``$b$" superscript; all our complexes will be bounded.

A \emph{chain map} $f$ from complex $\complex{E}$:
\begin{equation*}
 \cdots \xrightarrow{c_{n-1}} \sheaf{E}^n \xrightarrow{c_n}\sheaf{E}^{n+1}
\xrightarrow{c_{n+1}}\cdots
\end{equation*}
to complex $\complex{F}$:
\begin{equation*}
\cdots \xrightarrow{d_{n-1}} \sheaf{F}^n \xrightarrow{d_n}\sheaf{F}^{n+1}
\xrightarrow{d_{n+1}}\cdots
\end{equation*}
is a family of morphisms $f^n \in \mathrm{Hom}_\category{A}(\sheaf{E}^n, \sheaf{F}^n)$ satisfying
$f^{n+1} \circ c^n = d^n \circ f^n$. This definition of chain map  is nicely summarized  in the following commutative diagram:
\begin{equation}
\begin{CD}
\cdots @>{c_{n-1}}>>\sheaf{E}^n @>{c_n}>>\sheaf{E}^{n+1}
@>{c_{n+1}}>>\cdots\\
@. @V{f^n}VV @V{f^{n+1}}VV @. \\
\cdots @>{d_{n-1}}>>\sheaf{F}^n @>{d_n}>>\sheaf{F}^{n+1}
@>{d_{n+1}}>>\cdots
\end{CD}
\end{equation}

Before we move on to other induced categories, let's consider here some properties of complexes and chain maps.   For a given complex $\complex{E}: \cdots \xrightarrow{c_{n-1}} \sheaf{E}^n \xrightarrow{c_n}\sheaf{E}^{n+1}
\xrightarrow{c_{n+1}}\cdots$, one is tempted to define its cohomology as usual:
\begin{equation}\label{eq:CohoDef}
\mathrm{H}^n(\complex{E}) = \mathrm{Ker}\, c_n / \mathrm{Im}\, c_{n-1}
\end{equation}
For a general Abelian  category, $\category{A}$, a more subtle procedure is needed to define the cohomology, as the usual notion of \emph{``modding out''} the objects does not make sense in a general Abelian  category.
However, for the abelian categories we are interested in, the usual notions of kernels and cokernels make sense, and we can define the cohomology of the complex in the straightforward fashion \eqref{eq:CohoDef}.

Note that cohomology $\mathrm{H}^\bullet(\complex{E})$ itself can be regarded as a complex:
\begin{equation}
\cdots \xrightarrow{0} \mathrm{H}^n(\complex{E}) \xrightarrow{0} \mathrm{H}^{n+1}(\complex{E})
\xrightarrow{0}\cdots
\end{equation}
Then, one can easily verify that a chain map $f: \complex{E} \to \complex{F}$ induces another chain map $\mathrm{H}(f):
\mathrm{H}^\bullet(\complex{E}) \to \mathrm{H}^\bullet(\complex{F})$. If $\mathrm{H}(f)$ is an isomorphism, the chain map $f$ is  said to be a \emph{quasi-isomorphism}.

Here, let us take a moment to explain why these complicated mathematical concepts like complexes and quasi-isomorphisms play important roles in describing D-branes.  First, in 
\cite{Douglas:Categories,Diaconescu,Aspinwall-Lawrence:D0brane}, it is shown that a subclass of B-type D-branes on a Calabi-Yau $X$ -- namely those which correspond to branes in the B-twisted topological string theory -- can be described as complexes of coherent sheaves on $X$. It turns out that quasi-isomorphic complexes lead to identical open string spectra. So we would like to identify them as being ``isomorphic". But, in $\mathrm{Kom}(\category{A})$, quasi-isomorphisms are \emph{not} invertible. We need to pass to some fancied-up version, the \emph{derived category}, $D(\category{A})$, in which quasi-isomorphisms are turned into isomorphisms.
The objects of $D(\category{A})$ are complexes, as before. Just the morphisms have been changed.

But now an added payoff emerges, given a pair of complexes, $\CE^\bullet$ and $\CF^\bullet$, the space of morphisms in the derived category from $\CE^\bullet$ to $\CF^\bullet$, $\mathrm{Hom}_{D(\category{A})}(\CE^\bullet,\CF^\bullet)$ is isomorphic to the Hilbert space of states of the topological open string stretched between the corresponding D-branes (the space of \emph{chiral} operators in the corresponding physical open string). This is consistent with the above remark because isomorphic objects have the same (isomorphic) spaces of morphisms to (from) any other object. 

The procedure of inverting is similar to that of making abelian semi-groups into groups. For example, the number 4 does not have an inverse in the additive semi-group $\mathbb{Z}^{\ge 0} \equiv \mathbb{N} \cup \{0\}$. But, we can invert it by \emph{``creating''} its inverse $-4$. Inverting all positive integers in this way, we extend our semi-group to a group, $\mathbb{Z}$.

In the same way, we are going to extend our morphisms in $\mathrm{Kom}(\category{A})$ by creating inverses of all quasi-isomorphisms. Doing so, we will end up with a new category where quasi-isomorphisms in $\mathrm{Kom}(\category{A})$ have their inverses and hence are isomorphisms. This new category is called the \emph{derived category}. However, the procedure of inverting quasi-isomorphisms is not so simple as in 
our abelian semi-group example. This is because composing two morphisms is not commutative while the semi-group action was. To resolve this difficulty, we need to introduce, as an intermediate step, another category called the \emph{homotopy category},  $K(\category{A})$.

Objects of $K(\category{A})$ are the same as those of $\mathrm{Kom}(\category{A})$ but the morphisms are different. Two chain maps $f, f^\prime: \complex{E} \to \complex{F}$ are said to be \emph{homotopic} if there exists a family of morphisms $h^n \in \mathrm{Hom}_\category{A}(\sheaf{E}^n, \sheaf{F}^{n-1})$ such that $f^{\prime  n} = f^n + h^{n+1} \circ c^n + d^{n-1} \circ h^n$.
\begin{equation}
\begin{aligned}
\mbox{
\begindc[8]
    \obj(3,25){$\dots$}[Eldots]
    \obj(10,25){$\CE^{n-1}$}[Enm1]
    \obj(25,25){$\CE^{n}$}[En]
    \obj(40,25){$\CE^{n+1}$}[Enp1]
    \obj(47,25){$\dots$}[Erdots]
    \obj(3,15){$\dots$}[Fldots]
    \obj(10,15){$\CF^{n-1}$}[Fnm1]
    \obj(25,15){$\CF^{n}$}[Fn]
    \obj(40,15){$\CF^{n+1}$}[Fnp1]
    \obj(47,15){$\dots$}[Frdots]
    \mor{Eldots}{Enm1}{}
    \mor{Enm1}{En}{$c^{n-1}$}
    \mor{En}{Enp1}{$c^{n}$}
    \mor{Enp1}{Erdots}{}
    \mor{Fldots}{Fnm1}{}
    \mor{Fnm1}{Fn}{$d^{n-1}$}
    \mor{Fn}{Fnp1}{$d^{n}$}
    \mor{Fnp1}{Frdots}{}
    \mor(10,25)(10,15){$f^{'n-1}$}[\atleft,\dasharrow]
    \mor(9,25)(9,15){$f^{n-1}$}[\atright,\solidarrow]
    \mor(25,25)(25,15){$f^{'n}$}[\atleft,\dasharrow]
    \mor(24,25)(24,15){$f^{n}$}[\atright,\solidarrow]
    \mor(40,25)(40,15){$f^{'n+1}$}[\atleft,\dasharrow]
    \mor(39,25)(39,15){$f^{n+1}$}[\atright,\solidarrow]
    \mor(24,25){Fnm1}{$h^{n}$}[\atright,\solidarrow]
    \mor(39,25){Fn}{$h^{n+1}$}[\atright,\solidarrow]
\enddc
}
\end{aligned}
\end{equation}
The morphisms of $K(\category{A})$ are morphisms of $\mathrm{Kom}(\category{A})$ modulo homotopy equivalence. Since homotopic maps induce the same map on cohomology -- $\mathrm{H}(f')=\mathrm{H}(f)$ if $f$ is homotopic to $f'$ -- we have the same notion of quasi-isomorphisms as before. But
$K(\category{A})$ has a very special property regarding quasi-isomorphisms. Namely, the class of quasi-isomorphisms is \emph{localizing} in $K(\category{A})$. We will not give the rigorous definition of localization here (see \cite{Gelfand-Manin}). But we will make use of its consequences.

A \emph{roof} from complex $\complex{E}$ to complex $\complex{F}$ is a diagram $(s, f)$ of the form:
\begin{equation}
\begin{aligned}
\mbox{
\begindc[6]
    \obj(25,25){$\complex{G}$}[G]
    \obj(20,20){$\complex{E}$}[E]
    \obj(30,20){$\complex{F}$}[F]
    \mor{G}{E}{$s$}[\atright,\solidarrow]
    \mor{G}{F}{$f$}[\atleft,\solidarrow]
\enddc
}
\end{aligned}
\end{equation}
where $s$ is a quasi-isomorphism in $\mathrm{Hom}_{K(\category{A})}(\complex{G}, \complex{E})$ and
$f \in \mathrm{Hom}_{K(\category{A})}(\complex{G}, \complex{F})$. Two roofs are equivalent, $(s, f) \sim (t, g)$, if and only if there exists a third roof forming a commutative diagram of the form:
 \begin{equation}
\begin{aligned}
\mbox{
\begindc[6]
    \obj(25,25){$\complex{I}$}[I]
    \obj(20,18){$\complex{G}$}[G]
    \obj(30,18){$\complex{H}$}[H]
    \obj(15,11){$\complex{E}$}[E]
    \obj(35,11){$\complex{F}$}[F]	
    \mor{I}{G}{$r$}[\atright,\dasharrow]
    \mor{I}{H}{$h$}[\atleft,\dasharrow]
    \mor{G}{E}{$s$}[\atright,\solidarrow]
    \mor{G}{F}{$f$}[\atright,\solidarrow]
    \mor{H}{E}{$t$}[\atleft,\solidarrow]
    \mor{H}{F}{$g$}[\atleft,\solidarrow]
\enddc
}
\end{aligned}
\end{equation}
One of the nice consequences of localization is the following. Suppose we have two roofs, $(s, f): \complex{E} \to \complex{F}$ and $(t, g): \complex{F} \to \complex{G}$. Then there exists a third roof $(r, h)$ making the following diagram commutative:
 \begin{equation} \label{comp1}
\begin{aligned}
\mbox{
\begindc[6]
    \obj(25,25){$\complex{J}$}[J]
    \obj(20,18){$\complex{H}$}[H]
    \obj(30,18){$\complex{I}$}[I]
    \obj(15,11){$\complex{E}$}[E]
    \obj(25,11){$\complex{F}$}[F]
    \obj(35,11){$\complex{G}$}[G]	
    \mor{J}{H}{$r$}[\atright,\dasharrow]
    \mor{J}{I}{$h$}[\atleft,\dasharrow]
    \mor{H}{E}{$s$}[\atright,\solidarrow]
    \mor{H}{F}{$f$}[\atleft, \solidarrow]
    \mor{I}{F}{$t$}[\atright,\solidarrow]
    \mor{I}{G}{$g$}[\atleft,\solidarrow]
  \enddc
}
\end{aligned}
\end{equation}
Note that in the above diagram, $(s \circ r, g \circ h)$ is also a roof. This localizing property enables one to define the composition of two equivalence classes of roofs. That is
\begin{equation} \label{comp2}
[s, f] \circ [t, g] \equiv [s \circ r, g \circ h].
\end{equation}

Now we have all the technology to define the \emph{derived category} $D(\category{A})$.
\begin{itemize}
\item  Objects of $D(\category{A})$ are complexes in $\category{A}$.
\item Morphisms of $D(\category{A})$ are equivalence classes of roofs in 
$K(\category{A})$.
\item The composition law of morphisms is defined as in \eqref{comp1}, \eqref{comp2}.
\item The identity morphism $\mathit{id}_{\complex E}$ is $[\mathit{id}_{\complex E}, \mathit{id}_{\complex E}]$.
\end{itemize}
By defining morphisms of $D(\category{A})$ as above, we effectively \emph{``invert''} quasi-isomorphisms.
The inverse of quasi-isomorphism $[\mathit{id}, s]$ is $[s, \mathit{id}]$. Note $[\mathit{id}, s] \circ [s, \mathit{id}]
= [s, \mathit{id}] \circ [\mathit{id}, s] = [s, s] \sim [\mathit{id}, \mathit{id}]$ since the following diagram is commuting:
\begin{equation}
\begin{aligned}
\mbox{
\begindc[6]
    \obj(25,25){$\complex{E}$}[E1]
    \obj(20,18){$\complex{E}$}[E2]
    \obj(30,18){$\complex{F}$}[F1]
    \obj(15,11){$\complex{F}$}[F2]
    \obj(35,11){$\complex{F}$}[F3]	
    \mor{E1}{E2}{$\mathit{id}$}[\atright,\dasharrow]
    \mor{E1}{F1}{$s$}[\atleft,\dasharrow]
    \mor{E2}{F2}{$s$}[\atright,\solidarrow]
    \mor{E2}{F3}{$s$}[\atright,\solidarrow]
    \mor{F1}{F2}{$\mathit{id}$}[\atleft,\solidarrow]
    \mor{F1}{F3}{$\mathit{id}$}[\atleft,\solidarrow]
\enddc
}
\end{aligned}
\end{equation}

We had to pass to the homotopy category $K(\category{A})$ in order to define our roofs. If we'd tried to define them in the category of complexes, $Kom(\category{A})$, the composition of two roofs would not have been equivalent to another roof. Physically, imposing equivalence of chain maps up to homotopy is very closely related to the BRST construction of \cite{Diaconescu}.

There are fancier and more abstract definitions of the derived category
(see \cite{Gelfand-Manin}), but this one is \emph{constructive}, and allows you to actually compute the space of morphisms (and hence the spectrum of chiral operators of the open string theory) .

Before finishing this subsection, let's introduce some useful notations. As mentioned earlier, $\category{A}$ will be the category of coherent sheaves on a Calabi-Yau throughout the paper. Hence it will be useful to denote the (bounded) derived category of coherent sheaves on $X$ by $D(X)$. Also, we will frequently encounter complexes that consist of one entry. For example:
\begin{equation}
\cdots \to 0 \to \sheaf{E} \to 0 \to \cdots
\end{equation}
We will denote the above complex by $\sheaf{E}[n]$ where $n$ indicates $\sheaf{E}$ is at the $-n$ position in the complex. The position of the sheaves in the complex is usually called  the \emph{grading} in physics literatures. Sometimes, we will put small numbers  above the sheaves in a complex to denote their positions. For example:
\begin{equation}
\cdots \to \stackrel{-4}{\vphantom{\bigoplus_j}\sheaf{E}} \to \stackrel{-3}{\vphantom{\bigoplus_j}\sheaf{F}} \to \stackrel{-2}{\vphantom{\bigoplus_j}\sheaf{G}} \to \cdots
\end{equation}

Finally, there is an obvious functor from $D(X)$ to itself called the shift functor. It simply shifts the grading on all the objects (and morphisms). The complex $\complex{E}[n]$ denotes the complex $\complex{E}$ shifted $n$ places to the left.

\subsection{Object simplification}\label{sec:Object simplification}
Two complexes, $\CE_1^\bullet$ and $\CE_2^\bullet$, which are quasi-isomorphic lead to open string theories with the same spectrum (not just for the strings beginning and ending on $\CE_i^\bullet$, but also for the strings stretched between $\CE_i$ and and any other brane, $\CF^\bullet$). Thus we should identify $\CE_1^\bullet$ and $\CE_2^\bullet$ as representing the ``same" D-brane. We then might wish to ask to what extent we can use quasi-isomorphisms to replace a given complex by a ``simpler" one.

Consider the complex
\begin{equation}\label{eq:simplifyable}
  0\to\CE^0\xrightarrow{c_0}\CE^1
            \xrightarrow{c_1}\CE^2\to\dotsb\to \CE^{N-1}
            \xrightarrow{c_{N-1}}\CE^N\to 0
\end{equation}
Let us assume that \eqref{eq:simplifyable} is exact at the $\CE^0$ term ({\it i.e.}, that $c_0$ is injective). Then there is a quasi-isomorphism
\begin{equation}
\begin{CD}
  0 @>>> \CE^0 @>c_0>>     \CE^1    @>c_1>> \CE^2 @>c_2>> \dotsb\\
  @.       @.              @VqVV            @V1VV \\
     @.    0    @>>>\CE^1/c_0(\CE^0)@>c_1>> \CE^2 @>c_2>> \dotsb
\end{CD}
\end{equation}
So we can replace \eqref{eq:simplifyable} by the simpler complex \begin{equation*}\label{eq:simplifiedleft}
  0\to\CE^1/c_0(\CE^0)
            \xrightarrow{c_1}\CE^2\xrightarrow{c_2}\dotsb
\end{equation*}
By iterating this process, we can always eliminate all the exact terms on the left-hand side of the complex.

Conversely, consider the case where \eqref{eq:simplifyable} is exact at the $\CE^N$ term ({\it i.e.}~that $c_{N-1}$ is surjective). Then we can find a quasi-isomorphism
\begin{equation}
\begin{CD}
 \dotsb@>>>\CE^{N-2}@>c_{N-2}>> \CE^{N-1}  @>c_{N-1}>>\CE^N@>>>0\\
  @.         @A1AA                @AiAA  \\
 \dotsb@>>>\CE^{N-2}@>c_{N-2}>>\ker(c_{N-1})@>>>       0
\end{CD}
\end{equation}
Again, by iterating this process, we can eliminate all the exact terms on the right of the complex.

Combining the two operations, we can, without loss of generality, assume that there is cohomology in the first nonzero term in the complex and in the last nonzero term. If our complex is a direct sum of complexes, $\CE^\bullet=\CF_1^\bullet\oplus\CF_2^\bullet$, then we can apply this procedure to each of the direct summands separately.

But, in general, if the cohomology occurs in more than one term of the complex, we cannot simplify further. In particular, we typically \emph{cannot} reduce a complex of coherent sheaves, $\CE^\bullet$, to its cohomology for they are usually \emph{not} quasi-isomorphic.

As a simple (indeed, the prototypical) example of this, let $X=T^2\times T^2$, with coordinates $(z_1,z_2)$ and let the divisors $D_i=\{z_i=0\}$.

The complex
\begin{equation}\label{eq:d4d4bar}
    0\to \CO(-D_1)\xrightarrow{z_1}\CO\to 0
\end{equation}
has cohomology only in the second term and, indeed, is quasi-isomorphic to 
\begin{equation}\label{eq:d2}
    0\to 0\to\CO_{D_1}\to 0
\end{equation}
where $\CO_{D_1}$ is the structure sheaf of the divisor $D_1$ (extended by zero to a coherent sheaf on $X$). Physically, a wrapped D4-brane and a wrapped anti-D4 brane (which carries one unit of D2-brane charge on its worldvolume) can annihilate into a D2-brane wrapped on $D_1$ via tachyon condensation and 
\eqref{eq:d4d4bar} and \eqref{eq:d2} are different ways of expressing the endpoint of that condensation.

So far so good. But now consider the complex
\begin{equation}\label{eq:cannotsimplify}
    0\to \CO(-D_1)\oplus\CO(-D_2)\xrightarrow{\left(\begin{smallmatrix}z_1&z_2
\end{smallmatrix}\right)}\CO\to 0
\end{equation}
The cohomology of this complex is
\begin{equation}\label{eq:ifwecould}
    0\to \CO(-D_1-D_2)\to\CO_{p}\to 0
\end{equation}
where $p=D_1\cap D_2=\{z_1=z_2=0\}$. But \eqref{eq:cannotsimplify}
and \eqref{eq:ifwecould} are \emph{not} quasi-isomorphic.

\subsection{The intersection pairing}
In \cite{Distler-Brunner:Torsion,Distler-Brunner:TorsionII},  one of the 
guiding principles was that the monodromies acting on the K-theory should preserve the skew-symmetric bilinear pairing on $K(X)$, which, in the Calabi-Yau context, can be written as
\begin{equation}\label{eq:Kpairing}
   (v,w)= \int_X ch(v\otimes \overline{w}) Td(X)
\end{equation}
This pairing can be written for objects in the derived category as\footnote{For a pair of coherent sheaves, $\mathrm{Hom}_{D(X)}(\CF[0],\CE[i])=\mathrm{Ext}^i(\CF,\CE)$, so \eqref{eq:DerivedPairing} reduces to
\begin{equation*}
   (\CE[0],\CF[0])=\sum_i(-1)^i \dim \mathrm{Ext}^i(\CF,\CE)
\end{equation*}
}
\begin{equation}\label{eq:DerivedPairing}
   (\CE^\bullet,\CF^\bullet)= \sum_i (-1)^i
     \dim \mathrm{Hom}_{D(X)} (\CF^\bullet, \CE[i]^\bullet)
\end{equation}
The sum on $i$ is always a finite one because we are working with the \emph{bounded} derived category.
Any functor $F: D(X)\to D(X)$ will automatically preserve  \eqref{eq:DerivedPairing}, provided
\begin{enumerate}
\item[a)] $F$ is fully faithful\footnote{A functor $F: \category{A}\to\category{B}$ of additive categories is \emph{fully faithful} if, for any $X,Y\in Ob(\category{A})$,
\begin{equation*}
   F: \mathrm{Hom}_\category{A}(X,Y)\to \mathrm{Hom}_\category{B}(F(X),F(Y))
\end{equation*}
is an isomorphism.
}.
\item[b)] $F$ commutes, up to quasi-isomorphism, with the shift functor $[n]$.
\end{enumerate}
This will certainly hold for the monodromies defined by the kernels in \S\ref{sec:Monodromies},\S\ref{sec:Quintic}.

\subsection{Quantum effects}
We will end this ``review" section with some comments on how some of the information (specifically, the grading) in the derived category can be wiped out by quantum string effects (as we will argue in \S\ref{sec:NewCategory}, the grading is already only well-defined modulo 6).

Consider the pair of objects
\begin{equation*}
\CE^\bullet\
=\ 0\to\CO\to0\to\CO\xrightarrow{\phantom{{}^{\textstyle\oplus2}}}0\ \simeq\
0\to\CO\to\CO\xrightarrow{\left(\begin{smallmatrix}0\\ 1\end{smallmatrix}\right)}\CO^{\oplus2}\to0
\end{equation*}
and
\begin{equation*}
  \CF^\bullet\ =\
0\to0\to0\to\CO^{\oplus2}\to0\ \simeq\
0\to\CO\xrightarrow{1}\CO\xrightarrow{\phantom{\left(\begin{smallmatrix}0\\ 1\end{smallmatrix}\right)}}
\CO^{\oplus2}\to0
\end{equation*}
$\CE^\bullet$ and $\CF^\bullet$ are clearly not quasi-isomorphic (they're equal to their cohomologies, which are clearly not isomorphic).
The open string CFT on $\CF^\bullet$ has 4 chiral primaries (which give rise to a $U(2)$ gauge theory on its world-volume), whereas the open string CFT on $\CE^\bullet$ has only 2 (giving rise to a $U(1)\times U(1)$ gauge theory).

However, both of these branes can be viewed as the endpoints of tachyon condensation of the unstable brane
\begin{equation*}
  \CG^\bullet\ \simeq\
0\to\CO\to\CO\to\CO^{\oplus2}\to0
\end{equation*}
This has a $U(2)\times U(1)\times U(1)$ gauge theory and, at a typical point in the large-radius regime, can decay to \emph{either} $\CE^\bullet$ or $\CF^\bullet$ by condensing \emph{different} tachyons. 

At least in the large-radius regime, one expects that $\CE^\bullet$ is actually an excited state of $\CF^\bullet$. But there isn't an obvious tachyon that one can condense. One expects, instead, that the decay of 
$\CE^\bullet$ to $\CF^\bullet$ proceeds via barrier-penetration in the world-volume gauge theory. This is an effect not readily visible in the perturbative open-string theory.

More generally, consider a D-brane corresponding to some complex of coherent sheaves, at a \emph{particular point} in the K\"ahler moduli space. It is, in general, not easy to decide whether this brane is stable (BPS) at this point in the moduli space. One must check the $\pi$-stability criterion of \cite{Douglas:Categories,DoFiRoI} for every distinguished triangle in which this complex participates. This is not an easy task.

So, in the following, we will blithely talk about the monodromy mapping \emph{this} complex of coherent sheaves into \emph{that} complex. But we will not make any claim that the resulting complex corresponds to a \emph{stable} D-brane in the large-radius regime.

%\vfill\eject
\section{Monodromies}\label{sec:Monodromies}
\begin{floatingfigure}{1.5in}
   \mbox{\includegraphics[width=1in]{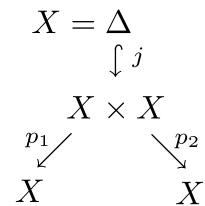}}
   \caption{Geometry associated to the monodromy action defined in \eqref{eq:autoequivdef}.}\label{fig:kernel}
\end{floatingfigure}
\noindent
Acting on the derived category, the monodromies are generated by kernels $K^\bullet\in D^b(X\times X)$. The formula for the monodromy action on a complex $\CE^\bullet$ is
\begin{equation}\label{eq:autoequivdef}
   \CE^\bullet\mapsto Rp_{1*}(K^\bullet\stackrel{L}{\otimes}p_2^*(\CE^\bullet))
\end{equation}
The name ``kernel" is deliberately chosen to remind you that this transformation is closely analogous to an integral transform,
\begin{equation*}
   f(x)\mapsto \int dy\ K(x,y)f(y)
\end{equation*}
Indeed, many of the formul\ae\ one writes for integral transforms ({\it e.g.}~for the composition of two such transforms) have precise analogues here. In \eqref{eq:autoequivdef}, we
\begin{enumerate}
\item Take a complex of sheaves $\CE^\bullet$ on $X$, ``pull it back" to the inverse-image complex of sheaves, $p_2^*(\CE^\bullet)$ on $X\times X$.
\item Take the tensor-product with the kernel, $K^\bullet$, and construct the left-derived complex of sheaves.
\item Finally, we ``push-forward" to the direct image complex, $p_{1*}(\cdot)$, and construct the right-derived complex of sheaves on $X$.
\end{enumerate}

Each of these steps sounds a little formidable, but, in practise, they are not. The left-derived functor, $\cdot \stackrel{L}{\otimes} \CF^\bullet$ is constructed by taking a complex, $V^\bullet$, of locally-free sheaves which is quasi-isomorphic to $\CF^\bullet$, and computing the \emph{ordinary} tensor-product with $V^\bullet$. So, in step 1, we replace $\CE^\bullet$ by a quasi-isomorphic complex of locally-free sheaves (sheaves of sections of holomorphic vector bundles $V^n$). The inverse-image of this complex is the complex of sheaves of sections of the pullback bundles $p_2^*V^n$. This is again locally-free, so we can take the ordinary tensor product with $K^\bullet$. Finally, in step 3, we need to take the direct-image. For our purposes, it will suffice to know the formula for the direct-image of two types of sheaves. One is the exterior tensor product of a sheaf on $X_1$ and a sheaf on $X_2$,
\begin{equation}
   R^ip_{1*}(\CE\boxtimes\CF)= \CE\otimes \coho{i}{X,\CF}
\end{equation}
The other case we will need is a sheaf supported on the diagonal $\Delta\stackrel{j}\hookrightarrow X\times X$ (extended by zero to a coherent sheaf on all of $X\times X$). This is 
\begin{equation}
\begin{split}
   R^ip_{1*}(\CE_\Delta)&= R^ip_{1*}j_*(\CE) =R^i(p_1\circ j)_* \CE\\
      &=\begin{cases}\CE&i=0\\0&\text{otherwise}\end{cases}
\end{split}
\end{equation}

\subsection{Some kernels}
The first, and most obvious monodromy is that about the large-radius limit (shifting the $B$-field by an integral class $\xi\in\coho{2}{X,\BZ}$). This acts on the D-branes by tensoring them with a line bundle $L$, with $c_1(L)=
\xi$. In terms of kernels,
\begin{equation}
   K_{r}^\bullet= (j_*L)[0]
\end{equation}
This action is a simple application of the  push-pull formula
\begin{equation*}
\begin{split}
   M_{r}(\CE^\bullet)&= Rp_{1*}((j_*L)[0] \stackrel{L}{\otimes}p_2^*(\CE^\bullet))\\&=
   Rp_{1*}(j_*( L\otimes j^*p_2^*(\CE^\bullet)))\\
   &= Rp_{1*}(j_*( L\otimes (p_2\circ j)^*(\CE^\bullet)))\\
   &= Rp_{1*}(j_*( L\otimes \CE)^\bullet)\\
   &= R(p_1\circ j)_*(( L\otimes \CE)^\bullet)\\
   &= ( L\otimes \CE)^\bullet
\end{split}
\end{equation*}
That is, we tensor the complex $\CE^\bullet$ term-by-term with the line bundle $L$.

The next, most obvious monodromy is that about the (mirror of the) conifold
(the principal component of the discriminant locus).  Kontsevich \cite{Kontsevich} conjectured that this was given by
\begin{equation}\label{eq:kontsevich}
   K^\bullet_c\stackrel{?}{=} 0\to
     \stackrel{-1}{\vphantom{\bigoplus_j}\CO_{X\times X}}\xrightarrow{r}
     \stackrel{ 0}{\vphantom{\bigoplus_j}\CO_\Delta}\to 0
\end{equation}
where $r$ is the restriction to the diagonal and where we have indicated by superscripts the grades of the sheaves in the complex. We will propose two modifications of this formula. First, we need to compose Kontsevich's kernel with the ``shift-by-two" functor.
\begin{equation}\label{eq:simplyconnconifold}
   K^\bullet_c= 0\to
     \stackrel{-3}{\vphantom{\bigoplus_j}\CO_{X\times X}}\xrightarrow{r}
     \stackrel{-2}{\vphantom{\bigoplus_j}\CO_\Delta}\to 0
\end{equation}
The reason is that we want $M_c(\CO[0])=\CO[0]$. Physically, the D6-brane becomes massless at the conifold locus, and there is a local description of the physics near the conifold locus when it is included as a fundamental field in the 4D action. Thus it must be single-valued with respect to $M_c$. With \eqref{eq:simplyconnconifold}, we get 
\begin{equation}\label{eq:simpleD6conifold}
\begin{split}
   M_c(\CO[0])&\simeq Rp_{1*}(0\to
     \stackrel{-3}{\vphantom{\bigoplus_j}\CO\boxtimes\CO}\xrightarrow{r}
     \stackrel{-2}{\vphantom{\bigoplus_j}\CO_\Delta}\to 0)\\
    &\simeq 0\to
     \stackrel{-3}{\vphantom{\bigoplus_j}\CO}\xrightarrow{1}
     \stackrel{-2}{\vphantom{\bigoplus_j}\CO}\to 0\to
     \stackrel{ 0}{\vphantom{\bigoplus_j}\CO}\to 0\\
    &\simeq \CO[0]
\end{split}
\end{equation}
%: yyy
where we used the fact that $\coho{0}{X,\CO}=\coho{3}{X,\CO}=\BC$. Had we used \eqref{eq:kontsevich} instead, we would have gotten $M_c(\CO[k])=\CO[k-2]$ \footnote{After completing this paper, it came to our attention that some of the monodromy calculations of this and the next section are done in \cite{Aspinwall-Douglas:Monodromy}, but using \eqref{eq:kontsevich} instead of \eqref{eq:simplyconnconifold}.}.

Our second modification is required to fix \eqref{eq:simplyconnconifold} on nonsimply-connected Calabi-Yau manifolds. If the holonomy of $X$ is $SU(3)$ and not a proper subgroup, then the fundamental group $\pi_1(X)$ is finite.
We can associate a flat holomorphic vector bundle, $W_j$ to each irreducible representation of $\pi_1(X)$. We replace \eqref{eq:simplyconnconifold} with
\begin{equation}\label{eq:conifoldkernel}
   K^\bullet_c= 0\to
     \stackrel{-3}{\bigoplus_j W_j\boxtimes W_j^*}\xrightarrow{\tilde r}
     \stackrel{-2}{\vphantom{\bigoplus_j}\CO_\Delta}\to 0
\end{equation}
where $\tilde r$ is restriction to the diagonal, followed by taking the trace. This sum over irreducible representations of $\pi_1(X)$  will induce the correct action on the K-theory, as computed in \cite{Distler-Brunner:TorsionII}, namely
\begin{equation}
   v\mapsto v-\sum_j(v,W_j)W_j
\end{equation}
where $v$ is the K-theory class of the D-brane in question, and $(\cdot,\cdot)$ is the skew-symmetric bilinear form \eqref{eq:Kpairing} on $K^0(X)$.

For the examples we will study, it is fairly easy to prove that the sheaf cohomology groups, $\coho{i}{X,W_j}=0$ except for the trivial representation, $W_j=\CO$. Similarly, we have
\begin{equation}
W_j^*\otimes W_k=\begin{cases}\CO\oplus\dotsb&j=k\\ \dotsb&\text{otherwise}\end{cases}
\end{equation}
where ``$\dotsb$" denote flat bundles associated to nontrivial irreps of $\pi_1(X)$. As a consequence,
\begin{equation}\label{eq:genD6conifold}
\begin{split}
   M_c(W_k[0])&\simeq Rp_{1*}\Bigl(
      K^\bullet_c\stackrel{L}{\otimes} p_2^*(W_k[0])\Bigr)\\
     &\simeq Rp_{1*}\Bigl(0\to
     \stackrel{-3}{\bigoplus_j W_j\boxtimes
        (W_j^*\otimes W_k)}\xrightarrow{\tilde r}
     \stackrel{-2}{\vphantom{\bigoplus_j}(W_k)_\Delta}\to 0\Bigr)\\
     &\simeq 0\to
     \stackrel{-3}{\vphantom{\bigoplus_j}W_k}\xrightarrow{1}
     \stackrel{-2}{\vphantom{\bigoplus_j}W_k}\to 0\to
     \stackrel{ 0}{\vphantom{\bigoplus_j}W_k}\to 0\\
    &\simeq W_k[0]
\end{split}
\end{equation}
This is what we expect, since, as we argued in \cite{Distler-Brunner:TorsionII}, all of the 6-branes corresponding to the $W_j$ become massless at the conifold. Hence they all must be single-valued under $M_c$.

For another example, let us return to the case of $X$ simply-connected and 
choose a smooth, irreducible effective divisor $D\in X$. Let $s$ be the holomorphic section of $\CO(D)$ whose divisor is $D$. Let us assume that the typical vanishing theorem, $\coho{i}{X,\CO(D)}=0$ for $i>0$, holds. Then a 4-brane wrapped on $D$ is represented by the sheaf $\CO_D$, which has the locally-free resolution
\begin{equation}
   \CO_D[0]\simeq  0\to
     \stackrel{-1}{\vphantom{\bigoplus_j}\CO(-D)}\xrightarrow{s}
     \stackrel{ 0}{\vphantom{\bigoplus_j}\CO}\to 0
\end{equation}
Let us compute what happens to this 4-brane when we circle the conifold.
\begin{equation*}
   K_c^\bullet\stackrel{L}{\otimes}p_2^*(\CO_D[0])\simeq
      0\to
     \stackrel{-4}{\vphantom{\bigoplus_j}\CO\boxtimes\CO(-D)}\xrightarrow{\left(\begin{smallmatrix}1\boxtimes s\\ r\end{smallmatrix}\right)}
     \stackrel{-3}{\vphantom{\bigoplus_j}\CO_{X\times X}\oplus \CO_\Delta(-D)}\xrightarrow{\left(\begin{smallmatrix}-r & s\end{smallmatrix}\right)}
     \stackrel{-2}{\vphantom{\bigoplus_j}\CO_\Delta}\to 0
\end{equation*}
Now, pushing down to $X$,
\begin{equation*}
\begin{split}
M_c(\CO_D[0])&\simeq
   Rp_{1*}\Bigl(K_c^\bullet\stackrel{L}{\otimes}p_2^*(\CO_D[0])\Bigr)\\
  &\simeq 0\to
     \stackrel{-3}{\vphantom{\bigoplus_j}\CO\oplus\CO(-D)}\xrightarrow{\left(\begin{smallmatrix}-1 & s\end{smallmatrix}\right)}
     \stackrel{-2}{\vphantom{\bigoplus_j}\CO} \xrightarrow{0}
     \stackrel{-1}{\vphantom{\bigoplus_j}\CO\otimes V}\to 0
\end{split}
\end{equation*}
or
\begin{equation}\label{eq:simpleD4conifold}
M_c(\CO_D[0])\simeq 0\to
     \stackrel{-3}{\vphantom{\bigoplus_j}\CO(-D)}\to 0\to
     \stackrel{-1}{\vphantom{\bigoplus_j}\CO\otimes V}\to 0
\end{equation}
where
\begin{equation}
   V= \ker\bigl(\coho{3}{X,\CO(-D)}\xrightarrow{s}\coho{3}{X,\CO}\bigr)
\end{equation}
So the wrapped D4-brane has turned into a collection of wrapped anti-D6-branes (one, at grade $-3$, with one unit of D4-brane charge dissolved on it; the rest at grade $-1$).

We can repeat the same calculation in the nonsimply-connected case, with the kernel \eqref{eq:conifoldkernel}. The result is
\begin{equation}\label{eq:conifoldD4gen}
  M_c(\CO_D[0])\simeq 0\to
     \stackrel{-3}{\vphantom{\bigoplus_j}\CO(-D)}\to 0\to
     \stackrel{-1}{\CO\otimes V_0\oplus{\bigoplus_j}' W_j\otimes V_j}\to 0
\end{equation}
where the direct sum $\bigoplus'_j$ runs over nontrivial irreps of $\pi_1(X)$ and
\begin{equation}\label{eq:Vjdef}
\begin{split}
  V_0&=\ker\bigl(\coho{3}{X,\CO(-D)}\xrightarrow{s}\coho{3}{X,\CO}\bigr)\\
  V_j&=\coho{3}{X,\CO(-D)\otimes W^*_j}
\end{split}
\end{equation}

As a final exercise (the details of which we will leave to the reader)
let us compute the monodromy about the conifold of one of those ``exotic" objects, discussed at the end of \S\ref{sec:Object simplification}, which is not isomorphic to its cohomology. Let $X$ be a simply-connected Calabi-Yau, and pick a pair of smooth irreducible effective divisors, $D_i$, which meet transversally. The complex
\begin{equation}
  \CE^\bullet\simeq0\to
\stackrel{-1}{\vphantom{\bigoplus_j}\CO(-D_1)\oplus\CO(-D_2)}
\xrightarrow{\left(\begin{smallmatrix}s_1&s_2\end{smallmatrix}\right)}
\stackrel{ 0}{\vphantom{\bigoplus_j}\CO}\to 0
\end{equation}
is not quasi-isomorphic to its cohomology
\begin{equation}
   \coho{\bullet}{\CE^\bullet}\simeq0\to
\stackrel{-1}{\vphantom{\bigoplus_j}\CO(-D_1-D_2)}\to
\stackrel{ 0}{\vphantom{\bigoplus_j}\CO_{D_1\cap D_2}}\to 0
\end{equation}
Going around the conifold, however, it turns into
\begin{equation}
   M_c(\CE^\bullet)\simeq 0\to
\stackrel{-3}{\vphantom{\bigoplus_j}\CO(-D_1)\oplus\CO(-D_2)}\to0\to
\stackrel{-1}{\vphantom{\bigoplus_j}\CO\otimes U}\to 0
\end{equation}
where
\begin{equation}
   U=\ker\left(\coho{3}{X,\CO(-D_1)}\oplus\coho{3}{X,\CO(-D_2)}
\xrightarrow{\left(\begin{smallmatrix}\hat s_1&\hat s_2\end{smallmatrix}\right)}\coho{3}{X,\CO}
\right)
\end{equation}
which is equal to its cohomology.

\section{The Quintic and its Orbifold}\label{sec:Quintic}
To obtain further concrete results, we need to specialize to some examples.
We will look at D-branes on $Y$, the quintic hypersurface in $\BP^4$, and on the nonsimply-connected Calabi-Yau manifold $X=Y/\BZ_5$, where we quotient the quintic by the freely-acting $\BZ_5$ symmetry (which exists for special choice  of defining polynomial) given by the action
\begin{equation}
   (x_1,x_2,x_3,x_4,x_5)\mapsto (\alpha x_1, \alpha^2 x_2,\alpha^3 x_3,
\alpha^4 x_4,x_5), \qquad \alpha^5=1
\end{equation}
on the homogeneous coordinates of $\BP^4$.

The line bundles on the quintic are entirely specified by their degree, and we denote them by $\CO(n)$. On $X$, this is no longer true. The divisors
$D_i=\{x_i=0\}$ and $D_j=\{x_j=0\}$ are no longer linearly-equivalent. Rather, there are flat, but nontrivial, line bundles
\begin{equation}
  \CL_{i-j}=\CO(D_i-D_j)
\end{equation}
Since $\BZ_5$ is abelian, all its irreducible representations are one-dimensional, and the associated vector bundles, $W_j$ are just the flat line bundles discussed above,
\begin{equation}
   W_j=\CL_j
\end{equation}

More generally, the line bundles of degree $n$ carry an extra label,
$\CO(n_\gamma)$, where $\gamma\in \BZ/5\BZ$. At degree-0, the flat line bundles, $\CL_j$, form a group under tensor product, of which the trivial bundle, $\CO$, is the identity element. When the degree $n\neq0$, however, there is no natural choice of ``origin" for the index $\gamma$.

The conformal field theory on $X$ can be viewed as an orbifold of the conformal field theory on $Y$. As such, there is a $\BZ_5$ quantum symmetry. This quantum symmetry acts on the D-branes (objects in the derived category) on $X$ by tensoring with the flat line bundles, $\CL_j$. The different choices of ``origin" for the index $\gamma$ are permuted by the action of the quantum symmetry group (the line bundles of degree $n$ form a module for the quantum symmetry group).

On the quintic, $Y$, the monodromy about the large-radius limit is given by tensoring with $\CO(1)$,
\begin{equation}
   M_r(\CE^\bullet)= (\CE\otimes\CO(1))^\bullet\equiv \CE(1)^\bullet
\end{equation}
On $X$, we need to \emph{choose} a particular line bundle of degree-one, $\CO(1_i)$,
\begin{equation}
   M_r(\CE^\bullet)=\CE(1_i)^\bullet
\end{equation}
Different choices of degree-one bundle are permuted by the quantum symmetry.

\subsection{Monodromies about the conifold point}
We have already written general formul\ae\ for the monodromy about the conifold of wrapped D6-branes \eqref{eq:simpleD6conifold},\eqref{eq:genD6conifold} and wrapped D4-branes \eqref{eq:simpleD4conifold},\eqref{eq:conifoldD4gen}. On the quintic , we have
\begin{equation}
\begin{split}
   M^Y_c(\CO[0])&\simeq\CO[0]\\
   M^Y_c(\CO_D[0])&\simeq 0\to
     \stackrel{-3}{\vphantom{\bigoplus_j}\CO(-1)}\to 0\to
     \stackrel{-1}{\vphantom{\bigoplus_j}\CO^{\oplus 4}}\to 0
\end{split}
\end{equation}
where $D$ is a hyperplane (a divisor of degree-one) in $Y$.
On $X$, we have
\begin{equation}
\begin{split}
   M^X_c(\CL_j[0])&\simeq\CL_j[0]\\
   M^X_c(\CO_{D_j}[0])&\simeq 0\to
     \stackrel{-3}{\vphantom{\bigoplus_j}\CO(-1_{5-j})}\to 0\to
     \stackrel{-1}{\vphantom{\bigoplus_j}\CO^{\oplus 4}}\to 0
\end{split}
\end{equation}
where $D_j$ is the hyperplane $\{x_j=0\}$ and we used the fact that the $V_k$ in \eqref{eq:Vjdef} are one-dimensional, $V_0=0$ and that
\begin{equation}
   \bigoplus_{j=1}^4\CL_j=\CO^{\oplus4}
\end{equation}

Now let's go further and consider higher codimension wrapped D-branes\footnote{We should point out an error in \cite{Distler-Brunner:Torsion,Distler-Brunner:TorsionII}. While the basis for the K-theory used there was perfectly fine for the topological considerations of those papers ({\it e.g.}, the computation of the action of the monodromies on the K-theory), we did not present a correct physical identification of the associated D-brane states. In particular, we wrote that the K-theory class $\CO(D)-\CO$ corresponds to a D4-brane wrapped on the divisor $D$. In fact, such a D4-brane corresponds to the K-theory class $\CO-\CO(-D)$. Also, the K-theory class that we said corresponded to a D2-brane wrapped on a curve $C$ is actually the \emph{complex-conjugate} of the K-theory class of such a wrapped D2-brane.
While these mis-identifications were irrelevant to the K-theoretic considerations of \cite{Distler-Brunner:Torsion,Distler-Brunner:TorsionII}, they \emph{are} relevant to our more ``refined" considerations here. For present purposes, we wish to use a basis of wrapped D-branes, rather any old basis of $K(X)$. In such a basis, for instance, the table on page 22 of \cite{Distler-Brunner:Torsion} would read
\begin{center}
\begin{tabular}{|c|c|c|c|c|c|}
\hline
&$r$&$c_1$&$c_2$&$c_3$\\
\hline\hline
$\CO$&$1$&$0$  & $0$ & $0$\\
$\CO_D=a-a^2+a^3$&$0$&
$\xi_2$  & $\xi_2^2$ & $\xi_2^3$\\
$\alpha=\CL-\CO$&0&$\chi$&$0$&$0$\\
$\CO_C=a^2-2a^3$&$0$&$0$  & $-\xi_2^2$ & $-2\xi_2^3$\\
$\CO_p=a^3$&$0$&
$0$  & $0$ & $2\xi_2^3$ \\
\hline
\end{tabular}
\end{center}
where we had previously used $a=\CO(D)-\CO$ and its tensor powers as a basis for $K(X)$.}. On $X$, the intersection of two hyperplanes is a rational curve,
\begin{equation}
   C_{jk}=\{x_j=x_k=0\}
\end{equation}
A 2-brane wrapped on $C_{jk}$ is represented by the sheaf $\CO_{C_{jk}}$, which has the locally-free resolution,
\begin{equation}
   \CO_{C_{jk}}[0]\simeq 0\to
   \stackrel{-2}{\vphantom{\bigoplus_j}\CO(-2_{5-j-k})}
   \xrightarrow{\left(\begin{smallmatrix}s_j\\ s_k\end{smallmatrix}\right)}
   \stackrel{-1}{\vphantom{\bigoplus_j}\CO(-1_{5-k})\oplus \CO(-1_{5-j})}
   \xrightarrow{\left(\begin{smallmatrix}-s_k& s_j\end{smallmatrix}\right)}
   \stackrel{ 0}{\vphantom{\bigoplus_j}\CO} \to 0
\end{equation}
For later convenience, we'll denote the maps above as $f=\left(\begin{smallmatrix}s_j\\ s_k\end{smallmatrix}\right)$ and $g=\left(\begin{smallmatrix}-s_k& s_j\end{smallmatrix}\right)$, with $g\circ f=0$ and the corresponding induced maps on sheaf cohomology by $\hat f$ and $\hat g$.

Following the familiar steps, we tensor this sequence with the kernel
\begin{equation*}
      K^\bullet_c= 0\to
     \stackrel{-3}{\bigoplus_{j=0}^4 \CL_j\boxtimes \CL_{5-j}}\xrightarrow{\tilde r}
     \stackrel{-2}{\vphantom{\bigoplus_j^4}\CO_\Delta}\to 0
\end{equation*}
to obtain
\begin{multline*}
      K^\bullet_c\stackrel{L}{\otimes}p_2^*(\CO_{C_{jk}}[0])\simeq
   0\to
   \stackrel{-5}{\bigoplus_{i=0}^4 \CL_i\boxtimes\CO(-2_{5-i-j-k})}\\
 \xrightarrow{\left(\begin{smallmatrix}f\\ \tilde r\end{smallmatrix}\right)}
     \stackrel{-4}{\bigoplus_{i=0}^4\CL_i\boxtimes
      (\CO(-1_{5-i-k})\oplus \CO(-1_{5-i-j}))\oplus\CO_\Delta(-2_{5-j-k})}\\
\xrightarrow{\left(\begin{smallmatrix}g &0\\ -\tilde r&
f\end{smallmatrix}\right)}
     \stackrel{-3}{\bigoplus_{i=0}^4\CL_i\boxtimes\CL_{5-i}\oplus
      \CO_\Delta(-1_{5-j})\oplus\CO_\Delta(-1_{5-j})}
\xrightarrow{\left(\begin{smallmatrix}\tilde r &g\end{smallmatrix}\right)}
\stackrel{-2}{\vphantom{\bigoplus_{i=0}^4}\CO_\Delta}
\to 0
\end{multline*}
Pushing forward, we get
\begin{multline*}
      Rp_{1*}(K^\bullet_c\stackrel{L}{\otimes}
      p_2^*(\CO_{C_{jk}}[0]))\simeq0\to
     \stackrel{-4}{\vphantom{\bigoplus_{i=0}^4}\CO(-2_{5-j-k})}
\xrightarrow{\left(\begin{smallmatrix}f\\0\end{smallmatrix}\right)}
     \stackrel{-3}{\vphantom{\bigoplus_{i=0}^4}
     \CO(-1_{5-j})\oplus\CO(-1_{5-j})\oplus\CO}\\
\xrightarrow{\left(\begin{smallmatrix}0&0\\ g&1\end{smallmatrix}\right)}
\stackrel{-2}{\bigoplus_{i=0}^4\CL_i\otimes \coho{3}{X,\CO(-2_{5-i-j-k})}\oplus\CO}\\
\xrightarrow{\left(\begin{smallmatrix}\hat f&0\\ 0&0\end{smallmatrix}\right)}
\stackrel{-1}{\bigoplus_{i=0}^4\CL_i\otimes (\coho{3}{X,\CO(-1_{5-i-k})}\oplus\coho{3}{X,\CO(-1_{5-i-j})})}
\xrightarrow{\hat g}
\stackrel{ 0}{\vphantom{\bigoplus_{i=0}^4}\CO}\to 0
\end{multline*}
This is quasi-isomorphic to
\begin{multline*}
      M^X_c(\CO_{C_{jk}}[0])\simeq
   0\to
     \stackrel{-3}{\vphantom{\bigoplus_{i=0}^4}
     \CI_{C_{jk}}}
\xrightarrow{0}
\stackrel{-2}{\bigoplus_{i=0}^4\CL_i\otimes \coho{3}{X,\CO(-2_{5-i-j-k})}}\\
\xrightarrow{\hat f}
\stackrel{-1}{\bigoplus_{i=0}^4\CL_i\otimes (\coho{3}{X,\CO(-1_{5-i-k})}\oplus\coho{3}{X,\CO(-1_{5-i-j})})}
\xrightarrow{\hat g}
\stackrel{ 0}{\vphantom{\bigoplus_{i=0}^4}\CO}\to 0
\end{multline*}
where $\CI_{C_{jk}}$ is the \emph{ideal sheaf} of the curve $C_{jk}$.
A short computation shows that the above sequence is exact everywhere except at the $-2$ position. So it is quasi-isomorphic to its cohomology,
\begin{equation*}
     M^X_c(\CO_{C_{jk}}[0])\simeq
   0\to
     \stackrel{-3}{\vphantom{\bigoplus_{i=0}^4}
     \CI_{C_{jk}}}
\xrightarrow{0}
\stackrel{-2}{\CO^{\oplus2}\oplus \bigoplus_{i=1}^4 \CL_i}\to 0
\end{equation*}
or, finally,
\begin{equation}
     M^X_c(\CO_{C_{jk}}[0])\simeq
   0\to
     \stackrel{-3}{\vphantom{\bigoplus_{i=0}}
     \CI_{C_{jk}}}
\xrightarrow{0}
\stackrel{-2}{\vphantom{\bigoplus_{i=0}}\CO^{\oplus6}}\to 0
\end{equation}

We can do a similar computation on the quintic. The intersection of two hyperplanes in $Y$ is, generically, a genus 6 curve, $C$. Under the conifold monodromy, a D2-brane wrapped on $C$ becomes
\begin{equation}
     M^Y_c(\CO_{C}[0])\simeq
   0\to
     \stackrel{-3}{\vphantom{\bigoplus_{i=0}}
     \CI_{C}}
\xrightarrow{0}
\stackrel{-2}{\vphantom{\bigoplus_{i=0}}\CO\otimes W}\to 0
\end{equation}
where
\begin{equation}
   W=\ker\Bigl(\coho{3}{Y,\CO(-2)}\xrightarrow{\left(\begin{smallmatrix}s\\ s'\end{smallmatrix}\right)} \coho{3}{Y,\CO(-1)}^{\oplus2}\Bigr)
\end{equation}
is also 6-dimensional.

The intersection of three hyperplanes in $X$ is a point, $p\in X$. So we have the locally-free resolution of a D0-brane located at $p$,
\begin{equation}
  \CO_p[0]\simeq 0\to
  \stackrel{-3}{\vphantom{\bigoplus_{i=0}^2}\CO(-3_{4})}\xrightarrow{f}
  \stackrel{-2}{\bigoplus_{i=0}^2\CO(-2_{i})}\xrightarrow{g}
  \stackrel{-1}{\bigoplus_{j=1}^3\CO(-1_{5-j})}\xrightarrow{h}
  \stackrel{ 0}{\vphantom{\bigoplus_{i=0}^2}\CO} \to 0
\end{equation}
where
\begin{equation}\label{eq:pointkoszulmaps}
f=\begin{pmatrix}s_1\\ s_2\\ s_3\end{pmatrix},\qquad g=\begin{pmatrix}0&s_3&-s_2\\ -s_3&0&s_1\\ s_2&-s_1&0\end{pmatrix},\qquad h=\begin{pmatrix}s_1&s_2&s_3\end{pmatrix}
\end{equation}
Following the same procedure as before, we obtain
\begin{multline*}
Rp_{1*}\Bigl(K_c^\bullet\stackrel{L}{\otimes}p_2^*\CO_p[0]\Bigr)\simeq 0\to
\stackrel{-3}{\CI_p\oplus\bigoplus_{j=0,\dotsc,4}\CL_j\otimes\coho{3}{X,\CO(-3_{4-j})}}
\xrightarrow{\left(\begin{smallmatrix}0&\hat f\end{smallmatrix}\right)}
\stackrel{-2}{\bigoplus_{\substack{j=0,\dotsc,4\\ k=0,1,2}}\CL_j\otimes\coho{3}{X,\CO(-2_{k-j})}}\\
\xrightarrow{\hat g}
\stackrel{-1}{\bigoplus_{\substack{j=0,\dotsc,4\\ k=1,2,3}}\CL_j\otimes\coho{3}{X,\CO(-1_{5-k-j})}}\xrightarrow{\hat h}
\stackrel{ 0}{\vphantom{\bigoplus_{i=0}}\CO} \to 0
\end{multline*}
where $\hat f,\hat g$ and $\hat h$ are the induced maps on cohomology. It is easy to see that this complex is exact except at the $-3$ position, where we have
\begin{equation*}
   M^X_c(\CO_p[0])\simeq 0\to \stackrel{-3}{\CI_p\oplus\bigoplus_{j=1}^4\CL_j}\to 0
\end{equation*}
or
\begin{equation}
   M^X_c(\CO_p[0])\simeq (\CI_p\oplus \CO^{\oplus4})[3]
\end{equation}

On the quintic, the intersection of three hyperplanes is a set of 5 points.
Most of the computation is actually simpler than the one on $X$.
One readily finds
\begin{equation}\label{eq:quinticconifoldD0}
M^Y_c(\CO_{\sum p_i}[0])\simeq
0\to
\stackrel{-3}{\vphantom{\bigotimes}\CI_{\sum p_i}\oplus\CO\otimes E_3}
\xrightarrow{\left(\begin{smallmatrix}0&\hat f\end{smallmatrix}\right)}
\stackrel{-2}{\vphantom{\bigotimes}\CO\otimes E_2}\xrightarrow{\hat g}
\stackrel{-1}{\vphantom{\bigotimes}\CO\otimes E_1}\xrightarrow{\hat h}
\stackrel{ 0}{\vphantom{\bigotimes}\CO}\to 0
\end{equation}
where
\begin{equation}\label{eq:EiDef}
\begin{split}
\begin{aligned}
   E_3&=\coho{3}{Y,\CO(-3)},          &\qquad \dim(E_3)=35\\
   E_2&=\coho{3}{Y,\CO(-2)}^{\oplus3},&\qquad \dim(E_2)=45\\
   E_1&=\coho{3}{Y,\CO(-1)}^{\oplus3},&\qquad \dim(E_1)=15
\end{aligned}
\end{split}
\end{equation}
and $\hat f,\hat g$ and $\hat h$ are the maps on cohomology induced from
\eqref{eq:pointkoszulmaps}. To proceed further, we need to compute the cohomology of the complex
\begin{equation}\label{eq:Eicomplex}
  0\to E_3\xrightarrow{\hat f}E_2\xrightarrow{\hat g}E_1\xrightarrow{\hat h}\BC\to0
\end{equation}
One readily finds that the kernel of $\hat f$ is 4-dimensional, which means that its image is 31-dimensional. The kernel of $\hat g$ is 33-dimensional, which means that its image is 12-dimensional. Finally, the image of $\hat h$ is 1-dimensional, so its kernel is 14-dimensional. Thus the \emph{cohomology} of the complex \eqref{eq:Eicomplex} is
\begin{equation}\label{eq:Eicohomology}
  0\to \BC^{\oplus4}\to\BC^{\oplus2}\to\BC^{\oplus2}\to0\to0
\end{equation}
Unlike coherent sheaves, complexes of vector spaces always split, and hence can always be reduced to their cohomology. So we can substitute this into \eqref{eq:quinticconifoldD0} and obtain
\begin{equation}\label{eq:quinticconifoldD0cohomology}
0\to
\stackrel{-3}{\vphantom{\bigotimes}\CI_{\sum p_i}\oplus\CO^{\oplus4}}\to
\stackrel{-2}{\vphantom{\bigotimes}\CO^{\oplus2}}\to
\stackrel{-1}{\vphantom{\bigotimes}\CO^{\oplus2}}\to 0
\end{equation}

Let us summarize the conifold monodromies that we have found.
On the quintic, $Y$, we found
\begin{equation}
\boxed{
\begin{aligned}
        \CO[0]&\mapsto \CO[0]\\
    \CO_{D}[0]&\mapsto 0\to
\stackrel{-3}{\vphantom{\bigotimes} \CO(-1)}\to 0\to
\stackrel{-1}{\vphantom{\bigotimes} \CO^{\oplus4}}\to 0\\
    \CO_{C}[0]&\mapsto 0\to
\stackrel{-3}{\vphantom{\bigotimes} \CI_{C}}\xrightarrow{0}
\stackrel{-2}{\vphantom{\bigotimes} \CO^{\oplus6}}\to 0\\
\CO_{\sum p_i}&\mapsto0\to
\stackrel{-3}{\vphantom{\bigotimes}\CI_{\sum p_i}\oplus\CO^{\oplus4}}\xrightarrow{0}
\stackrel{-2}{\vphantom{\bigotimes}\CO^{\oplus2}}\xrightarrow{0}
\stackrel{-1}{\vphantom{\bigotimes}\CO^{\oplus2}}\to 0
\end{aligned}
}
\end{equation}
where $D$ is a hyperplane divisor, $C$ is a genus-6 curve (the intersection of two hyperplanes), $\sum p_i$ is a set of 5 points (the intersection of three hyperplanes). On $X=Y/\BZ_5$, we had
\begin{equation}
\boxed{
\begin{aligned}
        \CL_j[0]&\mapsto \CL_j[0]\\
    \CO_{D_j}[0]&\mapsto 0\to
\stackrel{-3}{\vphantom{\bigotimes} \CO(-1_{5-j})}\to 0\to
\stackrel{-1}{\vphantom{\bigotimes} \CO^{\oplus4}}\to 0\\
    \CO_{C_{jk}}[0]&\mapsto 0\to
\stackrel{-3}{\vphantom{\bigotimes} \CI_{C_{jk}}}\xrightarrow{0}
\stackrel{-2}{\vphantom{\bigotimes} \CO^{\oplus6}}\to 0\\
   \CO_p[0]&\mapsto 0\to
\stackrel{-3}{\vphantom{\bigotimes} \CI_p\oplus \CO^{\oplus4}}\to 0
\end{aligned}
}
\end{equation}

Before closing this subsection, it is instructive to follow the D0-brane a second time around the conifold point of $X$. One readily finds
\begin{equation}
   (M^X_c)^2(\CO_p[0])\simeq 0\to
\stackrel{-5}{\vphantom{\bigoplus}\CI_p\oplus\CO^{\oplus4}}\to 0\to
\stackrel{-3}{\vphantom{\bigoplus}\CO^{\oplus5}}\to 0
\end{equation}
This expression is dramatically simpler than the one proposed by \cite{Aspinwall-Lawrence:D0brane} for the monodromy of the D0-brane going twice around the conifold point of the quintic. 

\subsection{Monodromies about the Landau-Ginsburg point}
The remaining distinguished point in the K\"ahler moduli space of the quintic or its orbifold is the Landau-Ginsburg point. The monodromy about that point must satisfy
\begin{equation}
   M_{LG}=M_c\circ M_r
\end{equation}
So, on the quintic, the corresponding kernel is
\begin{equation}
   K^{Y\bullet}_{LG}=0\to
\stackrel{-3}{\vphantom{\bigoplus}\CO\boxtimes\CO(1)}\xrightarrow{\tilde r}
\stackrel{-2}{\vphantom{\bigoplus}\CO_\Delta(1)}\to 0
\end{equation}
Let us compute the orbit of the D6-brane, $\CO[0]$ under the Landau-Ginsburg monodromy. Let $V=\coho{0}{Y,\CO(1)}$.
\begin{subequations}\label{eq:quinticLGD6}
\begin{equation}
   M_{LG}^Y(\CO[0])\simeq 0\to
\stackrel{-3}{\vphantom{\bigoplus}\CO\otimes V}\xrightarrow{ev}
\stackrel{-2}{\vphantom{\bigoplus}\CO(1)}\to 0
\end{equation}
Acting again, we get
\begin{equation}
\begin{split}
   (M_{LG}^Y)^2(\CO[0])&\simeq 0\to
\stackrel{-6}{\vphantom{\bigoplus}\CO\otimes V\otimes V}
\xrightarrow{\left(\begin{smallmatrix}sym\\ ev\end{smallmatrix}\right)}
\stackrel{-5}{\vphantom{\bigoplus}\CO\otimes\coho{0}{Y,\CO(2)}\oplus\CO(1)\otimes V}
\xrightarrow{\left(\begin{smallmatrix}-ev& ev\end{smallmatrix}\right)}
\stackrel{-4}{\vphantom{\bigoplus}\CO(2)}\to 0\\
&\simeq 0\to
\stackrel{-6}{\vphantom{\bigoplus}\CO\otimes \wedge^2 V}
\xrightarrow{ev}
\stackrel{-5}{\vphantom{\bigoplus}\CO(1)\otimes V}
\xrightarrow{ev}
\stackrel{-4}{\vphantom{\bigoplus}\CO(2)}\to 0
\end{split}
\end{equation}
In similar fashion, we find
\begin{align}
  (M_{LG}^Y)^3(\CO[0])&\simeq0\to
\stackrel{-9}{\vphantom{\bigoplus}\CO\otimes \wedge^3 V}\to
\stackrel{-8}{\vphantom{\bigoplus}\CO(1)\otimes \wedge^2 V}\to
\stackrel{-7}{\vphantom{\bigoplus}\CO(2)\otimes V}\to
\stackrel{-6}{\vphantom{\bigoplus}\CO(3)}\to 0\\
  (M_{LG}^Y)^4(\CO[0])&\simeq0\to
\stackrel{-12}{\vphantom{\bigoplus}\CO\otimes \wedge^4 V}\to
\stackrel{-11}{\vphantom{\bigoplus}\CO(1)\otimes \wedge^3 V}\to
\stackrel{-10}{\vphantom{\bigoplus}\CO(2)\otimes \wedge^2 V}\to
\stackrel{ -9}{\vphantom{\bigoplus}\CO(3)\otimes V}\to
\stackrel{ -8}{\vphantom{\bigoplus}\CO(4)}\to 0\\
  (M_{LG}^Y)^5(\CO[0])&\begin{gathered}[t]\simeq0\to
\stackrel{-15}{\vphantom{\bigoplus}\CO\otimes \wedge^5 V}\to
\stackrel{-14}{\vphantom{\bigoplus}\CO(1)\otimes \wedge^4 V}\to
\stackrel{-13}{\vphantom{\bigoplus}\CO(2)\otimes \wedge^3 V}\to
\stackrel{-12}{\vphantom{\bigoplus}\CO(3)\otimes \wedge^2 V\oplus\CO}\\ \to
\stackrel{-11}{\vphantom{\bigoplus}\CO(4)\otimes V}\to
\stackrel{-10}{\vphantom{\bigoplus}\CO(5)}\to 0
\end{gathered}
\end{align}
\end{subequations}
The apparent ``extra" factor of $\CO[12]$ in (\ref{eq:quinticLGD6}e) deserves explanation. The kernel of the map
\begin{equation*}
\stackrel{-12}{\vphantom{\bigoplus}\CO\otimes\coho{0}{Y,\CO(4)}
\otimes V}\to \stackrel{-11}{\vphantom{\bigoplus}\CO\otimes \coho{0}{Y,\CO(5)}}
\end{equation*}
contains not just the usual antisymmetrized piece, but an additional piece from the derivatives of the quintic defining equation. This yields the ``extra" factor of $\CO$ in (\ref{eq:quinticLGD6}e).

The complexes \eqref{eq:quinticLGD6} should be familiar to the cognoscenti. The bundle $T^*_{\BP^4}(1)$ fits into the short exact sequence
\begin{equation}
   0\to T^*_{\BP^4}(1) \to \CO\otimes V \to \CO(1)\to 0
\end{equation}
So (\ref{eq:quinticLGD6}a) is quasi-isomorphic to
\begin{subequations}
\begin{equation}
   M_{LG}^Y(\CO[0])\simeq T^*_{\BP^4}(1)[3]
\end{equation}
And, similarly, the other complexes in \eqref{eq:quinticLGD6} are resolutions of
\begin{align}
   (M_{LG}^Y)^2(\CO[0])&\simeq (\wedge^2 T^*_{\BP^4})(2)[6]\\
   (M_{LG}^Y)^3(\CO[0])&\simeq (\wedge^3 T^*_{\BP^4})(3)[9]\\
   (M_{LG}^Y)^4(\CO[0])&\simeq (\wedge^4 T^*_{\BP^4})(4)[12]\simeq\CO(-1)[12]\\
   (M_{LG}^Y)^5(\CO[0])&\simeq\CO[12]
\end{align}
\end{subequations}
While the shifts in grade may be unfamiliar, these are exactly the bundles conjectured in \cite{Mayr:McKayCorresp} (see also \cite{Douglas-Diaconescu}) to be the ``fractional branes" at the LG point of the quintic.

Completely analogous results holds for $X=Y/\BZ_5$. The kernel
corresponding to circling the Landau-Ginsburg point is
\begin{equation}
K^{X\bullet}_{LG}=0\to
\stackrel{-3}{\bigoplus_{j=0}^4\CL_j\boxtimes\CO(1_{i-j})}\xrightarrow{\tilde r}
\stackrel{-2}{\vphantom{\bigoplus^4}\CO_\Delta(1_i)}\to 0
\end{equation}
(Recall that the large-radius monodromy involved a choice of degree-one line bundle $\CO(1_i)$.) There's a rank-4 bundle, $F$ defined by the short exact sequence
\begin{equation}
   0\to F\to \bigoplus_{j=0}^4\CO(-1_j)\to\CO\to0
\end{equation}
and the Landau-Ginsburg monodromies are
\begin{equation}
\begin{split}
    M^X_{LG}    (\CO[0])&\simeq F(1_i)[3]\\
   (M^X_{LG})^2 (\CO[0])&\simeq (\wedge^2F)(2_{2i})[6]\\
   (M^X_{LG})^3 (\CO[0])&\simeq (\wedge^3F)(3_{3i})[9]\\
   (M^X_{LG})^4 (\CO[0])&\simeq (\wedge^4F)(4_{4i})[12]\simeq\CO(-1_{4i})[12]\\
   (M^X_{LG})^5 (\CO[0])&\simeq \CO[12]\end{split}
\end{equation}
Again, we find that $M_{LG}^5$ is the shift by 12 functor. Note that the dependence on the particular choice of degree-one bundle for the large-radius monodromy drops out when one takes the $5^{th}$ power of $M_{LG}$.

\subsection{A new category}\label{sec:NewCategory}
The Landau-Ginsburg monodromy \emph{should} satisfy $M_{LG}^5=1$. Instead, we have found $M_{LG}^5=[12]$, the shift-by-twelve functor. However shifting the grade by 6 is supposed to be a complete physical equivalence in the topological theory -- it is just spectral flow in the open-string channel\footnote{Spectral flow by 3 takes you from the NS back to the NS sector, but with the opposite value of $(-1)^F$. That is, it turns branes into anti-branes. Spectral flow by 6 takes you back to the sector you started in, with the \emph{same} value of
$(-1)^F$.} -- so perhaps we should be satisfied with this result.

But we cannot \emph{really} be satisfied if we wish to cling strictly to the derived category. 
\begin{equation*}
\mathrm{Hom}_{D(X)}(\CE^\bullet,\CF^\bullet)\neq \mathrm{Hom}_{D(X)}(\CE^\bullet[12],\CF^\bullet)
\end{equation*}
for general $\CF^\bullet$. So there is no obvious sense in which we should consider $\CE^\bullet$ and $\CE^\bullet[12]$ to be isomorphic.

However, there is a relatively simple modification of the derived category in which these are isomorphic. Let us define a new category, $\widetilde{D}(X)$, whose objects are again bounded complexes of coherent sheaves on a Calabi-Yau 3-fold, $X$, but whose morphisms are
\begin{equation}
  \mathrm{Hom}_{\widetilde{D}(X)}(\CE^\bullet,\CF^\bullet)=\bigoplus_{n=-\infty}^\infty
  \mathrm{Hom}_{D(X)}(\CE^\bullet,\CF^\bullet[6n]) 
\end{equation}
Since we are dealing with bounded complexes, the sum on $n$ receives only a finite number of nonzero contributions.

These morphisms compose in the obvious way, given the isomorphism \begin{equation*}
\mathrm{Hom}_{D(X)}(\CE^\bullet[k],\CF^\bullet[k])\simeq\mathrm{Hom}_{D(X)}(\CE^\bullet,\CF^\bullet)
\end{equation*}
 Namely, if $F\in\mathrm{Hom}_{D(X)}(\CE^\bullet,\CF^\bullet[6n_1])$ and
$G\in\mathrm{Hom}_{D(X)}(\CF^\bullet,\CG^\bullet[6n_2])$, then $G\circ F$ is the obvious element of $\mathrm{Hom}_{D(X)}(\CE^\bullet,\CG^\bullet[6(n_1+n_2)])$.

  We propose $\widetilde{D}(X)$ as the correct category of B-type topological open strings. This category takes account of the fact that shifting the grade by 6 should be a complete physical equivalence of the open string theory, whereas the original bounded derived category did not.

By \emph{construction}, now, $\CE^\bullet$ and $\CE^\bullet[12]$ are isomorphic in the category $\widetilde{D}(Y)$ and hence represent the same D-brane. Which is to say that, in this new category, $M_{LG}^5\simeq\Bid$. 

None of our calculations in this paper are modified by this new proposal, though the formula \eqref{eq:DerivedPairing} for the intersection-pairing can now be ``simplified"
\begin{equation}\label{eq:NewDerivedPairing}
\begin{split}
   (\CE^\bullet,\CF^\bullet)&= \sum_{i=-\infty}^\infty (-1)^i
     \dim \mathrm{Hom}_{D(X)} (\CF^\bullet, \CE[i]^\bullet)\\
      &=\sum_{i=-2}^{+3} (-1)^i
     \dim \mathrm{Hom}_{\widetilde{D}(X)} (\CF^\bullet, \CE[i]^\bullet)
\end{split}
\end{equation}

This new category also explains a long-standing puzzle \cite{Douglas:Categories} about the correspondence between topological open strings and physical open strings. As you vary the K\"ahler moduli the relative grading between two D-branes can shift. Douglas required the grading to be $\BR$-valued, so that, at a point where the branes are mutually-BPS, it is $\BZ$-valued, in accordance with the derived category.
This poses a puzzle because -- when one wishes to untwist and recover the physical open strings -- unitarity requires the charge of a chiral primary to lie in the range $0\leq q\leq3$.

In the original derived category, $\CE^\bullet$ and $\CE^\bullet[6]$ are not quasi-isomorphic, and hence represent \emph{distinct} topological D-branes. But in $\widetilde{D}(X)$, they \emph{are} isomorphic, and hence represent the \emph{same} topological D-brane\footnote{And $\CE^\bullet[3]\simeq\CE^\bullet[-3]$ is the corresponding anti-brane.}. Similarly, the charge of open string states, rather than being $\BR$-valued is only $\BR/6$-valued. We can always choose the charge of an open string state to lie in the range $(-3,3]$. In the physical theory, we interpret it as a chiral primary (if the charge is negative, we do a further spectral flow by 3 units, changing a brane to an anti-brane) whose charge is in the unitary range.

As explained in \cite{Douglas:Categories}, if we look at the open string string theory stretched between a pair of branes $(\CE^\bullet,\CF^\bullet)$ as we vary the K\"ahler moduli, at some point we come to a place where the charge of one of the chiral primaries falls outside of the unitary range. At this point, that brane configuration \emph{ceases to exist}\footnote{It was already unstable (non-BPS); it simply continued to make sense as an unstable brane configuration up to this point.}. In the present formulation, it is replaced by a \emph{new} (anti-)brane configuration whose open string CFT   is related by $N=2$ spectral flow to the previous one.

There is still a 1-1 correspondence between D-branes and \emph{isomorphism classes of objects} in $\widetilde{D}(X)$ and between chiral primaries and morphisms in $\widetilde{D}(X)$. There would be no such correspondence (as argued  by Douglas \cite{Douglas:Categories}), if we used the category $D(X)$. The point is that there are \emph{more} isomorphisms in $\widetilde{D}(X)$ and hence \emph{fewer} isomorphism classes. Objects which represented distinct topological D-branes in $D(X)$ are isomorphic in $\widetilde{D}(X)$ and hence represent the same D-brane. The cost, however, is that the correspondence (the twist required to go from the topological theory back to the physical one) jumps discontinuously as we move in the moduli space.

\section{Boundary Linear $\sigma$-Model}
The use of the Gauged Linear $\sigma$-Models with boundary to model B-type D-branes has been studied by \cite{Hellerman-McGreevy:Toolshed,Stanford:BLsM,Indians:BLsM,Indians:CoherentSheaves}.
The most natural application of this technology describes D-branes corresponding to 
complexes of direct sums of holomorphic line bundles. Our approach is closely related to that of \cite{Hellerman-McGreevy:Toolshed,Stanford:BLsM}. However, instead of introducing boundary superfields with deformed chiral constraints \cite{Hellerman-McGreevy:Toolshed}, we introduce a fermionic gauge shift symmetry for the chiral boundary bosons and fermions similar to \cite{Distler:Trieste94,Distler:LG}. This provides a convenient way to describe arbitrary complexes using standard boundary superfields.

\subsection{Boundary superspace}
The boundary of the GL$\sigma$M breaks half of the supercharges of the $N=(2,2)$ bulk theory. Therefore the superspace at the boundary reduces to $(t,\theta,\bar\theta)$ with the superspace derivatives
\begin{equation}
\begin{aligned}
   \uD&=\frac{\partial}{\partial\theta}-i\bar\theta\partial_0 \\
   \uDbar&=-\frac{\partial}{\partial\bar\theta}+i\theta\partial_0
\end{aligned}
\end{equation}
which have the anti-commutation relation $\{\uD,\uDbar\}=2i\partial_0$. Now, we introduce bosonic chiral boundary superfields $\bwp$ and fermionic chiral boundary superfields $\bbeta$ \cite{Stanford:BLsM}, which are constrained by $\uDbar\bwp=0$ and $\uDbar\bbeta=0$. In components these fields are given by
\begin{equation}
\begin{split}
   \bbeta&=\beta-\sqrt{2}\theta g-i\theta\bar\theta\partial_0\beta \\
   \bwp&=\wp+\sqrt{2}\theta\xi-i\theta\bar\theta\partial_0\wp
\end{split}
\end{equation}
where $\xi$ and $g$ are auxiliary fields. The bulk chiral fields reduce at the boundary to bosonic chiral boundary fields.

As in \cite{Distler:Trieste94} an unconstrained superfield, which has the expansion
\begin{equation}
   \Lambda=\lambda+\theta\bar a-\bar\theta b+\theta\bar\theta (\sqrt{2}\xi_\lambda-i\partial_0\lambda)
\end{equation}
is needed for the implementation of the gauge shift symmetry of the boundary chiral fields.

\subsection{Boundary Linear $\sigma$-Model}
To describe a (bounded) complex of direct sums of holomorphic line bundles
\begin{equation}\label{eq:linebundlecomplex}
  0\to\dotsb\to\bigoplus_i\CO(\Id{m_i}{2n-1})
   \xrightarrow{\mc{2n-1}}\bigoplus_i\CO(\Id{m_i}{2n})
   \xrightarrow{\mc{2n}}\bigoplus_i \CO(\Id{m_i}{2n+1})
   \xrightarrow{\mc{2n+1}}\dots \to 0
\end{equation}
in terms of the BL$\sigma$M, we introduce the boundary superfields given in Table~\ref{tab:BSMfields} and their conjugates. 

The entries of the matrices $\mc{k}$ are quasi-homogeneous polynomials in  the bulk chiral superfields, $\Phi$, restricted to the boundary. In addition to their bulk charges the fields also carry a boundary charge. The $\bbeta$s and $\bwp$s both have boundary charge $+1$. These charge assignments differ from those of \cite{Stanford:BLsM} where the boundary bosons have the opposite charge assignments. For ease of notation, we have assumed that the bulk GL$\sigma$M has a single $U(1)$. In the general case, of multiple $U(1)$'s (a complete-intersection Calabi-Yau in a general compact toric variety), ``$m$" in the above formul\ae\ is a multi-index representing the $U(1)$ charges.

\begin{table}[ht]
\begin{center}
\begin{tabular}{|c|c|c|c|}
   \hline
   Superfield  & Constraint & Bulk charge & Boundary charge \\
   \hline\hline
   $\Id{\bbeta_i}{2n}$, $i=1,\ldots,\dim\CE^{2n}$ & $\uDbar\Id{\bbeta_i}{2n}=0$   & $\Id{m_i}{2n}$ & $+1$ \\
   $\Id{\bwp_i}{2n+1}$, $i=1,\ldots,\dim\CE^{2n+1}$ & $\uDbar\Id{\bwp_i}{2n+1}=0$ & $\Id{m_i}{2n+1}$ & $+1$\\
   $\mc{k}(\Phi)_{ij}$ & $\uDbar(\mc{k})_{ij}=0$ & $\Id{m_i}{k+1}-\Id{m_j}{k}$ & $0$ \\
   $\Id{\Lambda}{k}$ (fermionic) & $\textrm{unconstrained}$       & $0$ & $0$ \\
%   $\Id{\Omega}{k}$  (fermionic) & $\uDbar\Id{\Omega}{k}=0$      & $0$ & $0$ \\
   $\Id{\bzeta}{k}$ (fermionic)      & $\uDbar\Id{\bzeta}{k}=0$      & $0$ & $0$ \\
   \hline
\end{tabular}
\caption{Boundary superfields of the BL$\sigma$M}
\label{tab:BSMfields}
\end{center}
\end{table}

As discussed in \cite{Distler:Trieste94} in the context of $N=(0,2)$ GL$\sigma$Ms, we introduce a gauge shift symmetry for the boundary superfields. Namely, for the fermions
\begin{subequations}
\begin{equation}
   \left( \begin{matrix} \Id{\bbeta}{2n} \\ \Id{\Lambda}{2n} \end{matrix} \right)
   \to \left( \begin{matrix} \Id{\bbeta}{2n}+\Id{\Omega}{2n}\mc{2n-1}\Id{\bwp}{2n-1} \\
                             \Id{\Lambda}{2n}-\Id{\Omega}{2n} \end{matrix} \right)   
\end{equation}
and for the bosons,
\begin{equation}
   \left( \begin{matrix} \Id{\bwp}{2n+1} \\ \Id{\Lambda}{2n+1} \end{matrix} \right)
   \to \left( \begin{matrix} \Id{\bwp}{2n+1}+\Id{\Omega}{2n+1}\mc{2n}\Id{\bbeta}{2n} \\
                             \Id{\Lambda}{2n+1}-\Id{\Omega}{2n+1} \end{matrix} \right)
\end{equation}
\end{subequations}
where $\Omega^{(k)}$ is a fermionic boundary superfield, obeying the chirality constraint, $\uDbar\Id{\Omega}{k}=0$, and having zero boundary and bulk charges.
Note that these gauge symmetries are compatible with the constraints which we have imposed on the fields in Table~\ref{tab:BSMfields}. However, for these shift symmetries to be commuting and independent, the composition $\mc{k+1}\circ\mc{k}$ must be zero for all $k$. This, in turn, is exactly the condition for the sequence \eqref{eq:linebundlecomplex} to be a complex.

Now, we want to construct a gauge invariant Lagrangian for the boundary chiral superfields. The gauge invariant kinetic term (we ignore, for the moment, the restrictions to the boundary of the bulk gauge fields; they will reappear in the next subsection) in the Lagrangian is given by
\begin{equation}\label{eq:LKin}
   \LKin=\frac{1}{2}\sum_n \int \ud^2\theta
   \left(\:
 \abs{\Id{\bbeta}{2n}+\Id{\Lambda}{2n}\mc{2n-1}\Id{\bwp}{2n-1}}^2
+\abs{\Id{\bwp}{2n+1}+\Id{\Lambda}{2n+1}\mc{2n}\Id{\bbeta}{2n}}^2 \:\right)
\end{equation}   
In addition to the kinetic term, we add a gauge invariant term involving $\Id{\Lambda}{k}$ and $\Id{\bzeta}{k}$ 
\begin{equation}
   \LMul=\sum_k \int \ud\theta\:\Id{\bzeta}{k} \left(
    \uDbar\Id{\Lambda}{k}-\Id{\kappa}{k} \right)+\hc
\end{equation}
where the $\Id{\kappa}{k}$ are constants.
The fermionic boundary superfields $\Id{\bzeta}{k}$ act as Lagrange multipliers, and in components we have
\begin{equation}\label{eq:Lagrangianmultcomp}
   \LMul=\sum_k \left(-\sqrt{2} \Id{g}{k}_\zeta (\Id{b}{k}-\Id{\kappa}{k})+
         \sqrt{2}\Id{\xi}{k}\Id{\zeta}{k}\right)+\hc
\end{equation}
Hence, these Lagrange multipliers constrain $\Id{b}{k}=\Id{\kappa}{k}$ and $\Id{\xi}{k}_\lambda=0$.

Due to the gauge shift symmetry and the assignment of bulk and boundary charges, one cannot write down a superpotential involving the superfields $\Id{\bbeta}{2n}$ and $\Id{\bwp}{2n+1}$.

A convenient gauge fixing condition turns out to be
\begin{equation}\label{eq:gaugefix}
   \uD\Id{\Lambda}{k} =0 \Rightarrow
     \begin{cases}
        \Id{\bar a}{k}=0 &\\ 
	\partial_0\Id{\lambda}{k}=-\frac{\sqrt{2}}{2}i\Id{\xi}{k} &
     \end{cases}
\end{equation}
which, again, is compatible with \eqref{eq:Lagrangianmultcomp}. Gauge-fixing, of course, complicates the supersymmetry transformation rules of the fields. The original transformations must be composed with a gauge shift transformation to preserve \eqref{eq:gaugefix}. Which is why is it good to start with a manifestly gauge invariant formulation.

After integrating out the auxiliary fields of \eqref{eq:LKin}, we obtain in components the gauge-fixed Lagrangian
\begin{equation}
\begin{split}
   \LKin=\sum_n & \left(\: i\IdbarT{\beta}{2n}\partial_0\Id{\beta}{2n}
                 +i\IdbarT{\wp}{2n+1}\partial_0\Id{\wp}{2n+1} \right.  \\
&+\frac{1}{2}\IdbarT{\beta}{2n}\left(\abs{\Id{b}{2n+1}}^2\:\mcbarT{2n}\mc{2n}+
\abs{\Id{b}{2n}}^2\:\mc{2n-1}\mcbarT{2n-1}\right)\Id{\beta}{2n} \\
&\left.+\frac{1}{2}\IdbarT{\wp}{2n+1}\left(\abs{\Id{b}{2n+2}}^2\:\mcbarT{2n+1}\mc{2n+1}+
\abs{\Id{b}{2n+1}}^2\:\mc{2n}\mcbarT{2n}\right)\Id{\wp}{2n+1}\:\right)\\
		&+(\textrm{interaction terms})
\end{split}			 
\end{equation}			
This contains the usual kinetic terms for the boundary fermions and bosons. Imposing the constraints that follow from \eqref{eq:Lagrangianmultcomp},
we replace $\Id{b}{k}$ by $\Id{\kappa}{k}$.  For nonvanishing $\Id{\kappa}{k}$, certain linear combinations of the boundary fields become massive. More precisely, the massless modes of the boundary fermions $\Id{\bbeta}{2n}$ are those that are in the kernel of both $\mc{2n+1}$ and $\mcbarT{2n}$. The fermions in the kernel of $\mcbarT{2n}$ are \emph{not} in the image of $\mc{2n}$. Thus the massless fermions $\Id{\bbeta}{2n}$ are exactly the fields that are given by the cohomology of the complex at grading $2n$. The same analysis holds for the massless boundary bosons.
In the infrared, the massive degrees of freedom can be neglected, and only the massless boundary fields (corresponding to the cohomology of the complex) remain.

At first sight this seems to be a slight puzzle. Previously, in section \S\ref{sec:Object simplification}, we argued that in general a complex \eqref{eq:linebundlecomplex} is \emph{not} quasi-isomorphic to its cohomology. How did we manage, then, to reduce to the cohomology here?

The reason is that we have only performed our analysis at a particular \emph{point} in the Calabi-Yau manifold. At a point, a complex of vector bundles reduces to a complex of \emph{vector spaces}. In the category of vector spaces, any complex is quasi-isomorphic to its cohomology. However, the bulk field $\phi$ is a quantum field to be integrated over. As we vary $\phi$, the kernels and cokernels of the $\mc{k}(\phi)$ vary and can even jump in dimension. In general, there's no uniform way to integrate out some of the boundary fields and reduce the complex \eqref{eq:linebundlecomplex} to its cohomology.

If the complex \eqref{eq:linebundlecomplex} is exact at one end \emph{everywhere} in the Calabi-Yau space, then according to section~\S\ref{sec:Object simplification} a simplification is possible. This arises in the BL$\sigma$M in the following way. Consider the case where \eqref{eq:simplifyable} is exact at the right-most term. In the BL$\sigma$M this gives masses to all boundary fields at grading $N$ (\emph{globally} on the Calabi-Yau) due to the surjectivity of $\mc{N-1}$. Similarly, if the map $\mc{0}$ is injective, so that the complex is exact at grading $0$, then we can (globally) give mass to the boundary fields at that grading.

We also give masses to some of the fields at grading $N-1$ (respectively, grading $1$). But the number of massless fields that remain can jump as we move about on the Calabi-Yau. They parametrize, not a direct sum of line bundles, nor even a vector bundle, but a coherent sheaf.

Nonetheless, if the complex \eqref{eq:linebundlecomplex} is exact at that term, we can repeat the process and (globally) integrate out those boundary fields as well. \emph{Exactly} as in \S\ref{sec:Object simplification}, the process can be continued until we hit the first term in the complex with nontrivial cohomology.

All of this assumed a particular (nonzero) value of the $\Id{\kappa}{k}$. Strictly speaking, different values of the $\Id{\kappa}{k}$ correspond to different boundary linear $\sigma$-models, but (at least for nonzero $\Id{\kappa}{k}$) they all flow to the same BCFT in the IR. Setting the $\Id{\kappa}{k}$ to zero recovers the original complex \eqref{eq:linebundlecomplex}, where all of the associated boundary fields are massless and survive in the IR. The statement of the equivalence of the 
original complex \eqref{eq:linebundlecomplex} and the simpler complex obtained by following the procedure of \S\ref{sec:Object simplification} is the statement of the smoothness of the $\Id{\kappa}{k}\to 0$ limit.

\subsection{Large radius monodromy}
In the previous discussion, we glossed over the fact that the boundary fields are charged under the bulk gauge symmetry. To incorporate the coupling of the boundary fields to (the restriction to the boundary of) the bulk gauge fields, we make the standard replacements
\begin{equation}
\begin{split}
   \Id{\bbeta}{2n}_i&\mapsto
     \Id{\tilde\bbeta}{2n}_i=e^{\Id{m}{2n}_i \bfield{V}}\Id{\bbeta}{2n}_i\\
    \Id{\bwp}{2n+1}_i&\mapsto
\Id{\tilde\bwp}{2n+1}_i=e^{\Id{m}{2n+1}_i \bfield{V}}\Id{\bwp}{2n+1}_i\\
    c_k(\Phi)_{ij}&\mapsto c_k(\tilde\Phi)_{ij}=e^{(\Id{m}{k+1}_i-\Id{m}{k}_i)\bfield{V}}c_k(\Phi)_{ij}\end{split}
\end{equation}
 in the Lagrangian \eqref{eq:LKin}, where $\bfield{V}$ is the bulk vector superfield restricted to the boundary. Furthermore, the $\Theta$-angle dependence of the boundary Lagrangian is given by
\begin{equation}
   \Lag_\Theta=\frac{\Theta}{2\pi}\int \ud^2\theta\:\bfield{V} =\frac{\Theta}{2\pi}\cdot \tilde v
\end{equation}
where
\begin{equation}
   \tilde v=v_0+\frac{\sqrt{2}}{2}(\sigma+\bar\sigma)
\end{equation}
The trick \cite{Stanford:BLsM,Hellerman-McGreevy:Toolshed}, now, is to note that 
 we  must include a charge projection at the boundary to ensure the right number of Chan-Paton degrees of freedom, namely that only one boundary field will be excited at a time. This is done by adding a Lagrange multiplier term of the form
\begin{equation}\label{eq:CPlagrangianmultiplier}
   \Delta S=\varsigma\left(j_\mathrm{Bdry}(0)-1\right)
\end{equation}
to the action. Since the boundary current,
\begin{equation}
   j_\mathrm{Bdry}=
           \left.\sum_n \left( \abs{\Id{\tilde\bbeta}{2n}+\Id{\Lambda}{2n}\mc{2n-1}\Id{\tilde\bwp}{2n-1}}^2
+\abs{\Id{\tilde\bwp}{2n+1}+\Id{\Lambda}{2n+1}\mc{2n}\Id{\tilde\bbeta}{2n}}^2
	   \:\right)\right|_{\theta=\bar\theta=0}
\end{equation} 
 is conserved on-shell, imposing this condition at one time, ensures that it holds everywhere along the boundary. We have therefore taken the Lagrange multiplier, $\varsigma$,
to be a c-number, rather than a field.

For the large radius monodromy, the relevant terms in the action  are the couplings to $\tilde v$.
\begin{equation}\label{eq:actionLRM}
   S=\int\ud t\:\tilde v \left(j_\mathrm{Bulk}+\frac{\Theta}{2\pi}\right)
    +\varsigma\left(j_\mathrm{Bdry}-1\right)
+(\textrm{$\tilde v$ independent terms})
\end{equation}
where 
\begin{equation}
\begin{split}
   j_\mathrm{Bulk}=
    \sum_{n,i} &\left( \Id{m}{2n}_i
\abs{\Id{\tilde\bbeta}{2n}_i+\Id{\Lambda}{2n}(\mc{2n-1}\Id{\tilde\bwp}{2n-1})_i}^2 \right.\\
&\left.\left.+\Id{m}{2n+1}_i\abs{\Id{\tilde\bwp}{2n+1}_i+\Id{\Lambda}{2n+1}(\mc{2n}\Id{\tilde\bbeta}{2n})_i}^2
     \:\right)\right|_{\theta=\bar\theta=0}
\end{split}	   
\end{equation} 
is the bulk charge current of the boundary fields.

Going around the large radius point in the moduli space, shifts the $\Theta$-angle by $2\pi$. This adds one unit to the current $j_\mathrm{Bulk}$. Because of \eqref{eq:CPlagrangianmultiplier}, the current $j_\mathrm{Bdry}$ is restricted to one, and therefore after having shifted $\Theta\rightarrow\Theta+2\pi$, we can rewrite the boundary action \eqref{eq:actionLRM} as \cite{Stanford:BLsM}
\begin{equation}
   S=\int\ud t\:\tilde v \left(j_\mathrm{Bulk}+j_\mathrm{Bdry}+\frac{\Theta}{2\pi}\right)
     +\varsigma\left(j_\mathrm{Bdry}-1\right)+(\textrm{$\tilde v$ independent terms})
\end{equation}
The effect (since the $\tilde\bbeta$s and $\tilde\bwp$s have boundary charge $+1$) is to shift $\Id{m}{k}_i\mapsto \Id{m}{k}_i+1$ for each boundary field.
This corresponds to tensoring all of the line bundles in \eqref{eq:linebundlecomplex} by $\CO(1)$.

The monodromies about the conifold and Landau-Ginsburg points, as computed above, are clearly more subtle. Unfortunately, computing them involves probing (along at least part of the path) a regime where the BL$\sigma$M is strongly-coupled. We hope to return to this, more subtle, analysis elsewhere.

\section*{Acknowledgements}
We would like to thank D.~Freed, S.~Hellerman, A.~Iqbal and A.~Kashani-Poor for helpful conversations. JD would like to thank the participants of the UT {\it Geometry and String Theory Seminar} for serving as a useful sounding board for some preliminary attempts at the ideas presented here. HJ gratefully acknowledges the support of a \emph{Studienstiftung des Deutchen Volkes} scholarship.

%%%%%%%%%%%%%%%%%%%%%%%%%%%%%%%%%%%%%%%%%%%%%%%%%%%%%%%%%%%%%%%%%%%%%
%\appendix

%\bigskip
%\section{Whatever}

\bibliography{derived}
\bibliographystyle{utphys}

\end{document}